\documentclass[12pt,a4paper]{article}

\usepackage{pslatex}
\usepackage{graphicx}
\usepackage{amsmath}
\usepackage{amssymb}

\usepackage{setspace}

\makeatletter

\@addtoreset{equation}{section}
\makeatother

\newcommand{\be}{\begin{eqnarray}}
\newcommand{\ee}{\end{eqnarray}}
\newcommand{\nn}{\nonumber}
\newcommand{\bn}{\begin{enumerate}}
\newcommand{\en}{\end{enumerate}}

\def\IC{\mathbb{C}}

\def\IR{\mathbb{R}}
\def\IZ{\mathbb{Z}}

\def\CM{{\cal M}}
\def\CN{{\cal N}}
\def\CO{{\cal O}}

\def\CQ{{\cal Q}}

\def\a{\alpha}
\def\b{\beta}

\def\w{\omega}
\def\G{\Gamma}

\def\S{\Sigma}

\def\goto{\rightarrow}

\def\Tr{{\rm Tr}}

\newcommand{\bdC}{\mathbb{C}}
\newcommand{\mathcalQ}{\mathcal{Q}}

\begin{document}

\begin{titlepage}

\begin{flushright}
arXiv:0705.3540 
\\
SNUST-070501
\end{flushright}

\vspace*{2.0cm}
\centerline{\Large\bf M2-brane Probe Dynamics and Toric Duality}

\vspace*{2.0cm}

\centerline{
Seok Kim$^1$, Sangmin Lee$^2$, Sungjay Lee$^2$ and Jaemo Park$^{3,4}$}


\begin{center}
\sl $^1$ 
School of Physics, Korea Institute for Advanced Study, Seoul 130-722, Korea 
\\ 
$^2$ 
School of Physics and Astronomy, Seoul National University, 
Seoul 151-747, Korea
{}\\
$^3$ 
Department of Physics, Postech, 
Pohang 790-784, Korea
{}\\
$^4$
Postech Center for Theoretical Physics (PCTP),
Pohang, 790-784, Korea
\end{center}

\vspace{1.5cm}

\abstract{
We study the dynamics of a single M2-brane 
probing toric Calabi-Yau four-fold singularity 
in the context of the recently proposed M-theory crystal model 
of AdS$_4$/CFT$_3$ dual pairs. 
We obtain an effective abelian gauge theory in which
the charges of the matter fields are given by 
the intersection numbers between loops and faces in the crystal.  
We argue that the probe theory captures certain aspects 
of the CFT$_3$ even though the true M2-brane 
CFT is unlikely to be a usual gauge theory.
In particular, the moduli space of vacua of the gauge theory
coincides precisely with the Calabi-Yau singularity. 
Toric duality, partial resolution, and a possibility of new RG flows 
are also discussed.
}


\end{titlepage}

\section{Introduction}

Throughout the development of the AdS/CFT correspondence \cite{adscft} 
for the past decade, the AdS$_4$/CFT$_3$ cases in M-theory 
has remained much less understood than the AdS$_5$/CFT$_4$ cases in 
IIB string theory. 
The lack of a perturbative description of M-theory and its M2/M5-brane 
world-volume theories poses a major obstacle. 

When supergravity serves as a good approximation to the full theory, 
M-theory on AdS$_4$ is no more difficult than 
IIB string theory on AdS$_5$. 
The difficulty still remains on the CFT side. 
While the CFT$_4$ can be derived by quantizing open strings 
on the D3-branes probing a conical singularity, 
no systematic method is available to derive the world-volume 
theory of M2-branes near a singularity.
In early works on AdS$_4$/CFT$_3$, proposals for the CFT$_3$ were made 
based on global symmetries and 
analogy with quiver gauge theories on D-branes; 
see, for example, \cite{ot, ak, dall, fab, ahn1}. 
None of the proposals was systematic enough to be applied to 
less symmetric geometries such as the infinite families $Y^{p,q}$ 
\cite{gmsw1, gmsw2, clpp}. 

For M2-branes probing {\em toric} Calabi-Yau four-fold (CY$_4$) cones, 
a first step toward constructing the CFT$_3$ 
was taken in \cite{crystal1, crystal2}.
It is an M-theory generalization of the brane-tiling model \cite{t1, t2, t3, t4, t5},
which has proven extremely successful in studying the CY$_3$/CFT$_4$ counterpart. 
As the information on the CFT$_3$ is encoded 
in a three dimensional periodic lattice, 
the model was named the {\bf M-theory crystal model}. 
Following \cite{crystal2}, we will distinguish the D3/CY$_3$ model 
from the M2/CY$_4$ model by referring to the 
former as the tiling model and the latter as the crystal model. 

By construction, the tiling/crystal models share many features, for example,
\begin{center}
\begin{tabular}{rcl} 
Tiling/Crystal &  & CFT$_4$/CFT$_3$  \cr \hline
edges (bonds) & $\leftrightarrow$ & matter fields \cr
vertices (atoms) & $\leftrightarrow$ & super-potential terms \cr
faces (tiles) & $\leftrightarrow$ & gauge groups 
\end{tabular}
\end{center}
This is by now a well established dictionary in the tiling model. 
As for the crystal model, the first two items were verified in 
\cite{crystal1, crystal2}, but the third one posed a puzzle. 
The graph of the tiling model 
splits the unit cell ($T^2$) into disjoint faces (tiles), 
and the D5/NS5-brane interpretation \cite{t2,ima1} of the model explains 
why gauge groups are assigned to the faces. 
In contrast, the graph of the crystal model does not 
divide the unit cell ($T^3$) into disjoint cells. 
At first sight, the crystal model does not seem to 
assign any role to faces and cells.

There is another related problem. 
In the crystal model, the fundamental fields of the CFT$_3$ 
are interpreted as open M2-brane discs ending on M5-branes, 
and the ``gauge-invariant'' operators as closed spherical M2-branes 
formed by gluing the M2-discs. It requires a new algebraic 
structure beyond the usual matrix product. 
So we are tempted to conclude that the CFT$_3$ 
cannot be written down until  
the full ``non-abelian'' M5-brane theory is constructed,  
which is a notoriously difficult subject on its own.

If the ``non-abelian'' nature of M-branes is the major obstacle 
to further progress, we may ask whether the problem can be simplified 
by considering a single M2-brane probe.
In the D3/CY$_3$ case, the abelian gauge theory for a single D3-brane 
probe proved useful in the early stages of the development 
\cite{dgm, mp, he00} as well as in the discovery 
of the tiling model. In general, the probe theory does {\em not} 
exhibit the dynamics of the full non-abelian theory. 
For example, the super-potential often vanish in the abelian theory 
while it plays an essential role in the non-abelian theory. 
On the other hand, the probe theory {\em does} encode some important 
features of the full theory such as the multiplicity of matter super-fields 
and the moduli space of vacua which coincide with the CY$_3$ 
singularity. 

Bearing in mind both the usefulness and the limitation of the D3-brane probe theory, 
in the present paper, we explore the probe theory of an M2-brane 
in the context of the crystal model. The crystal model maps the M2-brane 
probe to an M5-brane. The information on the toric CY$_4$ geometry 
is reflected in the shape of the M5-brane in the ``internal'' space. 
We take the world-volume theory of the M5-brane to be a free CFT 
with a single tensor multiplet of $(2,0)$ supersymmetry in six dimension. 
Once we ``compactify'' the M5-brane theory along the internal space, 
we obtain an effective abelian gauge theory in $(2+1)$-dimensions 
with $\CN=2$ supersymmetry. 
As the open M2-brane discs are charged under the self-dual tensor field 
on the M5-brane, they indeed become charged matter fields in the probe theory.
Moreover, the gauge groups {\em can} be attributed to faces of the crystal 
as in the dictionary above. 
The charges of the matter fields are then given by 
the intersection numbers between loops and faces in the crystal. 

Some cautionary remarks are in order. 
There is no reason to expect that the M2-brane theory 
is a gauge theory. In fact, the crystal model points strongly to the contrary. 
The abelian gauge theory under discussion 
should be regarded, at best, as an approximate description of the true theory, 
to the same extent as the M5-brane theory is an abelian tensor gauge theory. 
Physical consequences derived from the probe theory 
may or may not survive in the full ``non-abelian'' theory. 
In either case, we find it reasonable to expect the probe theory to be a 
useful guide toward the construction of the ``non-abelian'' theory.

Partial resolution, which we study in section 6, is a good example to illustrate the point. 
We will show that the M2-brane probe theory admits partial Higgsing 
which translates to partial resolution of the CY$_4$ singularity. 
An important result here is that this theory does {\em not} 
allow all possible resolution that a D3-brane probe does. 
Suppose a dual explanation of this difference can be found in 
terms of the difference between IIB supergravity and M supergravity 
(a question we hope to address in a future work). 
Then, it would impose a non-trival constraint on any future 
candidates for the ``non-abelian'' CFT$_3$. 
Our discussion in section 7 on possible RG flows could serve similar purpose. 

This paper is organized as follows.
We begin with a brief review of the crystal model \cite{crystal1, crystal2} 
in section 2.
In section 3, we explain how to identify the $U(1)$ gauge groups and 
compute the charges from the crystal model. 
We also explain how the charge assignment rule can be derived from 
the world-volume theory of a single M5-brane. 
In section 4, we show that the moduli space of vacua 
coincides with the toric CY$_4$, 
thereby confirming the validity of the theory.
In section 5, we show that two different gauge theories can correspond to 
the same CY$_4$ in a way similar to the {\bf toric duality} 
of \cite{he00}. Toric duality of non-abelian CFT$_4$ 
is known to be a Seiberg duality. The meaning of toric duality of ``non-abelian'' 
CFT$_3$ is an interesting open question.  
In section 6, we study partial Higgsing of a theory to obtain another one, 
thereby partially resolving the CY$_4$ singularity, 
following similar consideration of the abelian CFT$_4$ \cite{he00}. 
In section 7, we note that the abelian theory allows a 
``massive deformation'' analogous to the Klebanov-Witten flow \cite{kw}
from an orbifold CFT to the conifold CFT. 
We conclude with some discussions in section 8.

\section{M-theory Crystal Model: a brief review}

\begin{figure}[htbp]
\begin{center}
\includegraphics[width=12cm]{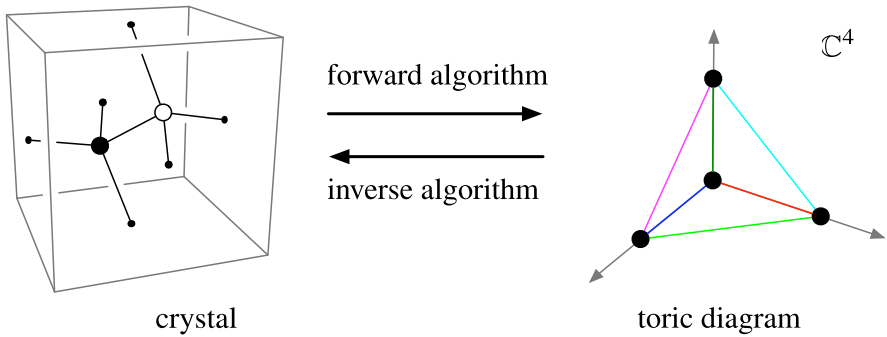}
\caption{Crystal vs. Toric diagram.} \label{fwdinv}
\end{center}
\end{figure}

We give a brief review of the M-theory crystal model \cite{crystal1, crystal2}. 
The crystal model relates a toric CY$_4$ to a three-dimensional periodic 
graph (a crystal) which encodes informations on the CFT$_3$. 
As in \cite{he00, t4, crystal2}, the maps between a crystal and the 
corresponding toric diagram will be called the forward and inverse algorithms; 
see Figure \ref{fwdinv}. 

The toric diagram forms a convex polyhedron in $\IZ^3 \subset \IZ^4$. 
The reduction from $\IZ^4$ to $\IZ^3$ is a consequence of the CY condition.
See, for example, \cite{msy1, msy2} for more information on toric geometry 
in this context.

The crystal model follows from a T-duality of M-theory. 
We take the T-duality transformation along $T^3\subset T^4$ 
in alignment with the projection $\IZ^4 \goto \IZ^3$. 
This corresponds to $x^{6,7,8}$ directions in Table 1 below.
By T-duality, we mean the element $t$ in the $SL(2,\IZ)\times SL(3,\IZ)$ duality 
group which acts as $t: \tau \equiv C_{(3)} + i \sqrt{g_{T^3}} \goto -1/\tau$.
The stack of $N$ M2-branes turns into a stack of $N$ M5-branes wrapping 
the dual $T^3$. 
We call them the $T$-branes. 
The degenerating circle fibers turn into another M5-brane 
extended along the (2+1)d world-volume and a non-trivial 3-manifold 
$\S$ in $\IR^3 \times T^3$. We call it the $\S$-brane. 
Preservation of supersymmetry requires that the $\S$-brane
wrap a special Lagrangian submanifold of  
$\IR^3 \times T^3 = (\IC^*)^3$, 
and that it is locally a plane in $\IR^3$ and a 1-cycle in $T^3$. 
The result is summarized in Table 1.
\begin{table}[htbp]
\label{brane3}
$$
\begin{array}{l|ccc|cccccc|cc}
\hline
           & 0 & 1 & 2 & 3 & 4 & 5 & 6 & 7 & 8 & 9 & 11\\
\hline
\mbox{M5} & \circ & \circ & \circ & & & & \circ & \circ & \circ \\
\mbox{M5} & \circ & \circ & \circ &  \multicolumn{6}{c|}{\Sigma} & & \\
\hline
\end{array}
$$
\caption{The brane configuration for the CFT$_3$.  
Away from the origin of $\IR^3$(345), 
the special Lagrangian manifold $\Sigma$ is locally a product of 
a 2-plane in $\IR^3$(345) and a 1-cycle in $T^3$(678).}
\end{table}

The crystal graph is  
the intersection locus between 
the $T$-branes and the $\S$-brane projected onto the $T^3$ 
(the {\bf alga} projection). 
It was shown in \cite{crystal1} that 
the matter fields of the CFT$_3$ are the 
M2-discs whose boundaries encircle the bonds of the crystal, 
hence the first item of the dictionary we discussed in the introduction. 
The derivation of the M2-disc picture made use of the fact that the $T$-branes 
and the $\S$-brane can be merged into a single, smooth, brane configuration. 
This observation will play a crucial role when we later 
discuss the gauge groups. 
It was also argued in \cite{crystal1, crystal2} that the atoms (vertices) of the 
crystal give the super-potential terms. As an evidence, it was shown 
to reproduce the BPS spectrum of meson operators of the CFT$_3$. 

In principle, the forward algorithm consists of two simple steps:
(a) reading off the CFT$_3$ (b) showing that its moduli space 
of vacua gives the toric CY$_4$. 
These steps were not directly verified in \cite{crystal1, crystal2}
since the information on the gauge group was missing. 
Instead, it was shown that the ``fast'' forward algorithm \cite{t1, t2} 
of the tiling model can be applied without any modification.
The fast forward algorithm is based on the concept of perfect matchings.
A perfect matching is a subset of bonds of the crystal, such that every atom 
of the crystal is an end-point of precisely one such bond. The bonds in each 
perfect matching carry an orientation. We choose to orient the arrows to go 
from a white atom to a black one. 
Perfect matchings have several nice properties. In particular, 
\bn
\item 
Each perfect matching can be located in the toric diagram. 
The relative coordinate in the toric diagram 
between two perfect matchings $p_\a$ and $p_\b$ 
is given by the homology charge of $(p_\a-p_\b)$ regarded 
as a one-cycle in $T^3$. 
\item 
The perfect matchings solve the `abelian' version of the F-term condition 
for the chiral fields $X_i$ associated to the bonds, if we set 
\be
\label{Fsol}
X_i = \prod_\a p_\a^{\langle X_i, p_\a \rangle}, \nn
\ee
where $\langle X_i, p_\a \rangle$ equals 1 if $p_\a$ contains the bond $X_i$ and 0 
otherwise. 
\en
The abelian gauge theory of the present paper fills the gap by showing that 
the ``slow'' forward algorithm also works. 
The perfect matchings will continue to be a useful device.

\begin{figure}[htbp]
\begin{center}
\includegraphics[height=3.5cm]{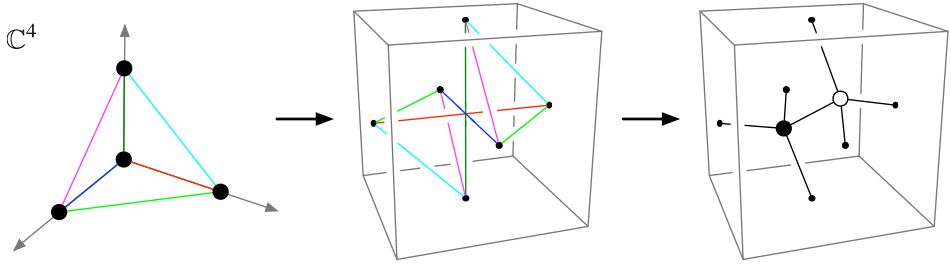}
\caption{Inverse algorithm for $\IC^4$, reproduced from \cite{crystal2}.} 
\label{c4inverse}
\end{center}
\end{figure}

The inverse algorithm is not as well established as the forward algorithm. 
Roughly speaking, one draws a 1-cycle in the $T^3$ for each edge of 
the toric diagram and then merge the 1-cycles to make up the crystal. 
For simple examples with high degree of symmetry, the procedure 
is unambiguous. Even in the general case, the result can be checked 
by using the forward algorithm. It is certainly desirable to understand 
the inverse algorithm better by finding an explicit expression 
for the special Lagrangian manifold $\S$.

\begin{figure}[htbp]
\begin{center}
\includegraphics[height=4cm]{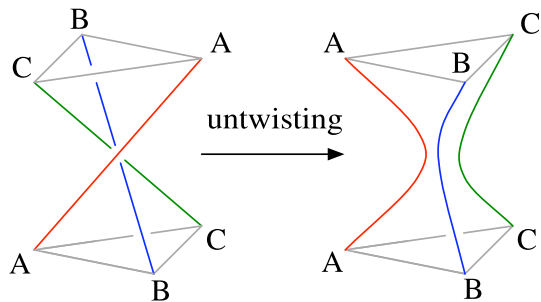}
\caption{Untwisting from alga to amoeba, reproduced from \cite{crystal2}.} 
\label{untwist}
\end{center}
\end{figure}

It is useful to consider the projection of the crystal onto $\S$ (the 
{\bf amoeba} projection).
We can think of the amoeba as a 3-manifold with some defects. 
As we shrink the 2-fans along the radial direction, the ``points at infinity'' form a locally one dimensional defect. The 1-cycles paired with the 2-fans are localized along the defects. Globally, the defect is isomorphic to the dual toric diagram. When the toric diagram has no internal points, the amoeba has the topology of 
a three-sphere apart from the defect.

\begin{figure}[htbp]
\begin{center}
\includegraphics[height=3.5cm]{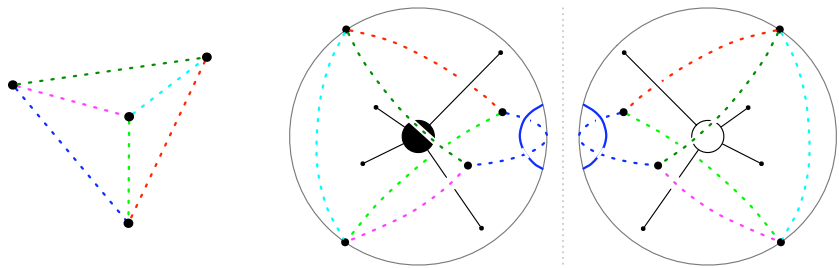}
\caption{Dual toric diagram and amoeba diagram for $\IC^4$, 
reproduced from \cite{crystal2}.
We represent a three-sphere as the union of two balls with the surfaces 
identified. The dotted lines on the balls denote the defects.} \label{c4amoeba}
\end{center}
\end{figure}

It is possible to obtain the amoeba from the alga and vice versa 
using the {\bf untwisting} procedure. The untwisting flips the orientation 
of the space transverse to the bond of the crystal
See Figure \ref{untwist}. We apply the untwisting 
map to the alga of $\IC^4$, we obtain the amoeba depicted in Figure \ref{c4amoeba}. 
Note that the dual 
toric diagram is a tetrahedron as expected. The 1-cycles are localized along the dual toric diagram as they should be. Applying the untwisting map to 
$C(Q^{1,1,1})$ yields a similar 
result, with the dual toric diagram being a cube.

\begin{figure}[htbp]
\begin{center}
\includegraphics[height=18cm]{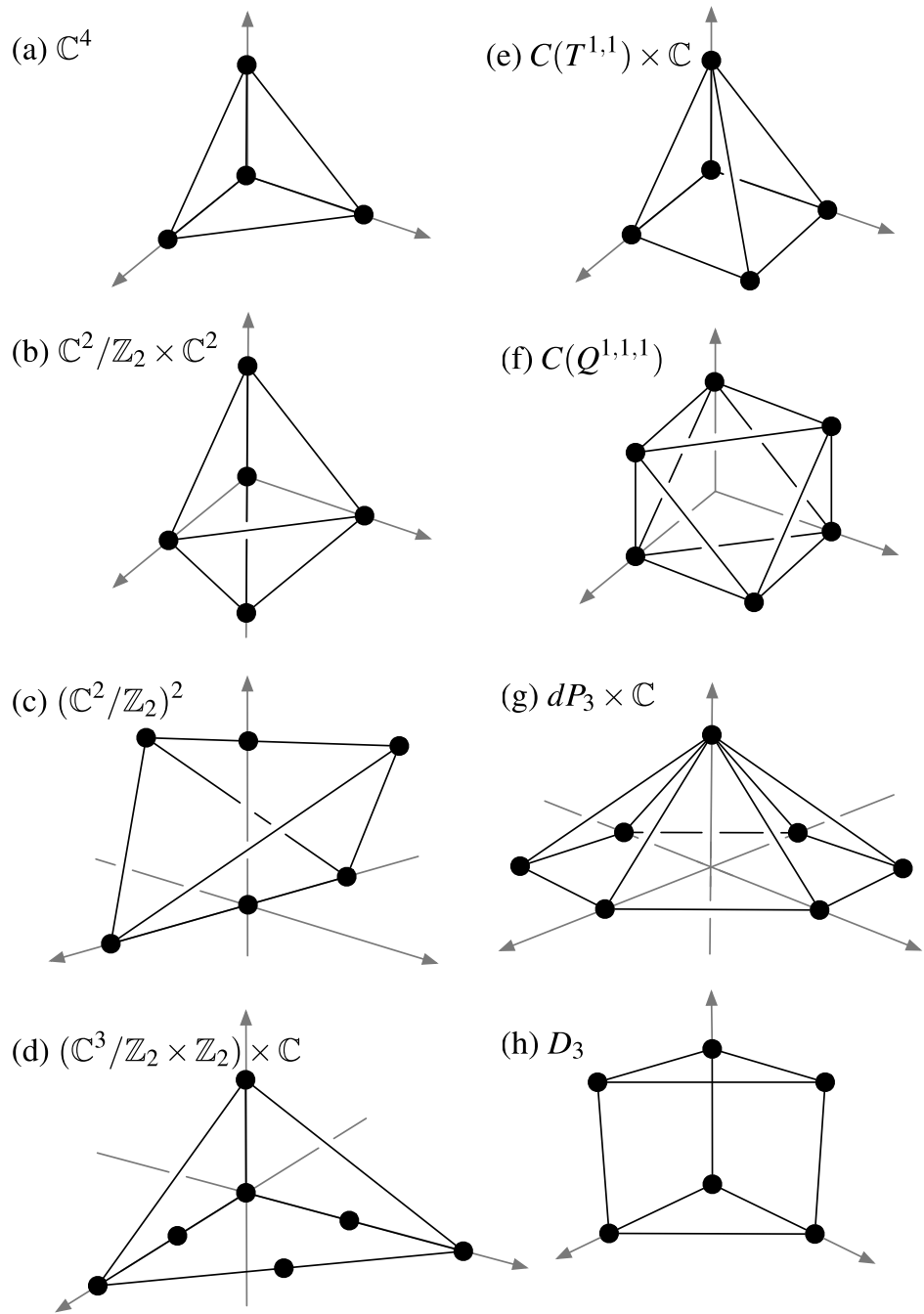}
\caption{Toric diagrams of the CY$_4$ considered in this paper.} \label{toric_all}
\end{center}
\end{figure}

We close this section with Figure \ref{toric_all} which contains 
the toric diagrams of CY$_4$ we consider in this paper. 
The corresponding crystals will be shown in later sections. 

\section{Abelian Gauge Theory}

\subsection{Charge assignment rule} 

\begin{figure}[htbp]
\begin{center}
  \hskip 2cm 
  \begin{minipage}[b]{0.45\linewidth}
     \includegraphics[width=4.0cm]{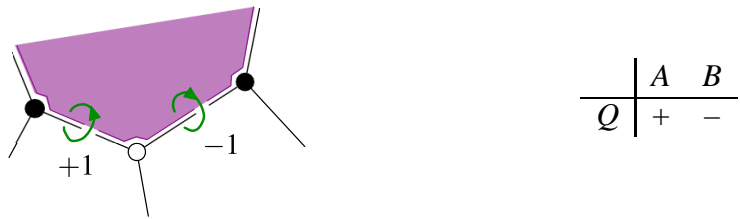}
   \end{minipage}
\begin{minipage}[b]{0.35\linewidth}
\raisebox{1.5cm}{
\begin{tabular}{r|cc}
  & $A$ & $B$  \cr \hline
$Q$ & + & --  \cr
\end{tabular}
}
  \end{minipage}
\caption{Charges as intersection numbers between the ``matter loops'' 
and the ``charge faces.''} \label{faceloop}
\end{center}
\end{figure}

The main feature of the abelian gauge theory 
associated to the crystal model 
is that the $U(1)$ gauge groups are attributed to faces of the crystal. 
Recall that a matter field in the crystal model 
is represented by an M2-brane disc whose boundary loop is localized 
along a bond of the crystal. 
Then, the charges of the matter fields are 
simply the intersection number between the ``matter loops'' 
and the ``charge faces.'' 
Figure \ref{faceloop} gives a pictorial description of the charge assignment rule.
In the figure, we fixed the orientation of the matter loops 
by imagining arrows from white atoms to black ones 
and applying the right-hand rule. This ensures that 
the M2-discs surrounding an atom can form an orientable two-sphere.

Although this is quite reminiscent of the charge assignment in the tiling model, 
we should emphasize the difference between the two models.
In the D5/NS5-brane picture of the tiling model, it is very clear 
how the disjoint tiles of D5-branes give rise to gauge groups 
for the matter fields that are open strings connecting the tiles.
The role of faces in the crystal model is not very clear at first sight. 
In addition, the number of independent gauge groups is not so easy to count 
as in the tiling model. 

\begin{figure}[htbp]
\begin{center}
\includegraphics[width=10.0cm]{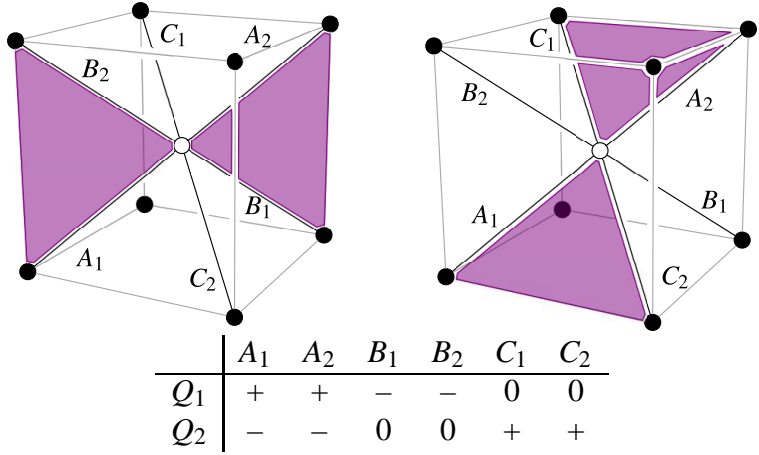}
\\
\begin{tabular}{r|cccccc}
  & $A_1$ & $A_2$ & $B_1$ & $B_2$ & $C_1$ & $C_2$ \cr \hline
$Q_1$ & + & + & -- & -- & 0 & 0 \cr
$Q_2$ & -- & -- & 0 & 0 & + & + 
\end{tabular}
\caption{Two independent faces for $C(Q^{1,1,1})$.} \label{q_ch}
\end{center}
\end{figure}

We will shortly explain the M-theory origin of the charge assignment as 
well as how to count the number of gauge groups. Before doing so, 
we give a concrete example of the charge assignment in Figure \ref{q_ch}. 
It has two independent charges. 
This example is especially easy to visualize because the faces lie on planes, 
which is not true in general. 

\subsection{M-theory origin} 

As we reviewed in section 2, the crystal model comes from 
two types of M5-branes: the $T$-brane and the $\S$-brane. 
An important step, which was used already in identifying the matter fields 
as open M2-branes, is that the two M5-branes can merge into a single 
M5-brane of a complicated topology. Let us call it $\CM$.
This M5-brane carries on its world-volume the  ``self-dual'' two-form field $B$.
The abelian gauge fields of the crystal model 
come from the Kaluza-Klein reduction of the $B$-field 
along the internal manifold $\CM$.

Open M2 branes ending on an M5-brane couples to the `self-dual' 
two-form field $B$ as
\be
S_{M2} \sim  \int_{\partial D} B  
\ee
When the M5-brane is extended over $\IR^{1,2}\times \CM$, 
we can use harmonic one-forms on $\CM$ to decompose $B$ as 
\be
B(x,y) = A^a(x) \wedge \w_a(y) + \cdots, 
\ee
where the omitted terms are required to ensure the self-duality condition 
$dB = * dB$.

An M2-disc whose boundary is a 1-cycle $C_i$ in $\CM$ become 
a particle in (2+1)-dimension with a coupling to the gauge field, 
\be
S_i \sim Q_{ia} \int A^a .
\ee
The charge is easy to read off
\be
Q_{ia} = \int_{C_i} \w_a = \sharp(C_i, S_a) . 
\ee
In the last step, we replaced the integral of $\w_a$ over $C_i$ 
by the intersection number between $C_i$ and the two-cycle $S_a$ 
which is Poincar\'e dual to $w_a$.

\begin{figure}[htbp]
\begin{center}
\includegraphics[width=10.0cm]{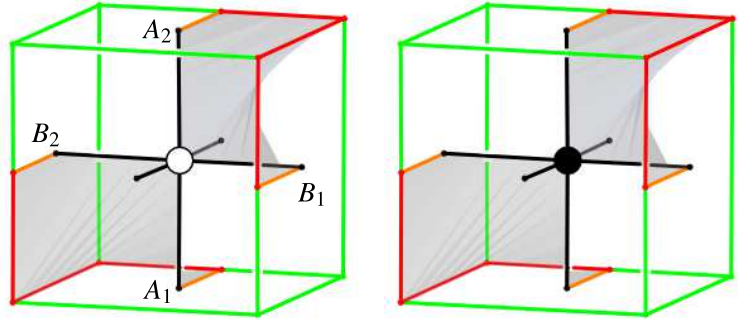}
\caption{A 2-cycle in the amoeba diagram of $C(Q^{1,1,1})$. 
The union of the two solid cubes with the surfaces identified 
represent a three-sphere. The edges of the cube (green line) 
denote the dual toric diagram at asymptotic infinity.
} \label{Q111-2cy}
\end{center}
\end{figure}

It remains to show that faces of the crystal define 2-cycles of $\CM$. 
When the $T$-brane and $\S$-brane merge to form $\CM$, 
the bonds of the crystal serve as the ``throat'' connecting the two components. 
So, the faces should be carried over to the $\S$ side. 
Recall that the transition from the crystal to the amoeba involves 
the untwisting process. It means that the tangent plane to the face 
go through a 180$^\circ$ rotation as it moves from 
an atom to its neighbor along a bond.

In the amoeba diagram, the bonds of the crystal and the dual toric diagram 
are linked non-trivially. So, it is impossible 
for the extended face to form a compact 2-cycle of $\CM$. 
The only alternative is that it runs off to infinity to form a non-compact 2-cycle. 
An example of a such a 2-cycle is depicted in Figure \ref{Q111-2cy}.
In the figure, the face emanates from the bonds 
($A_1$-$B_2$-$A_2$-$B_1$) where $A_i, B_i$ are define in Figure \ref{q_ch}. 
It is twisted in accordance with the untwisting process, and 
it runs off to the dual toric diagram at infinity. 
The red segments of the dual toric diagram denote the asymptotic boundary 
of the face. 

We have checked for all examples considered in this paper that 
any face of the crystal can be extended to the amoeba diagram and sent 
to asymptotic infinity. We take it as a strong evidence that it works 
for any crystal, but we have not found a general proof yet.

\subsection{Examples}

To illustrate the charge assignment rule, we present a few more examples in 
the following set of figures along with tables summarizing the charges. 
The faces are not drawn in the figures explicitly, but 
the charge table can be used to figure out the faces for each charge. 

\begin{figure}[htbp]
\begin{center}
\begin{minipage}[b]{0.35\linewidth}
\includegraphics[width=3.8cm]{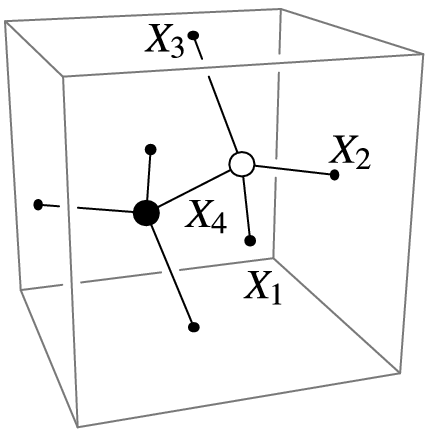}
\end{minipage}
\begin{minipage}[b]{0.35\linewidth}
\raisebox{2cm}{
All $X_i$'s are neutral and $W=0$.
}
\end{minipage}
\caption{$\IC^4$} \label{C4ch}
\end{center}
\begin{center}
\begin{minipage}[b]{0.35\linewidth}
\includegraphics[width=4.0cm]{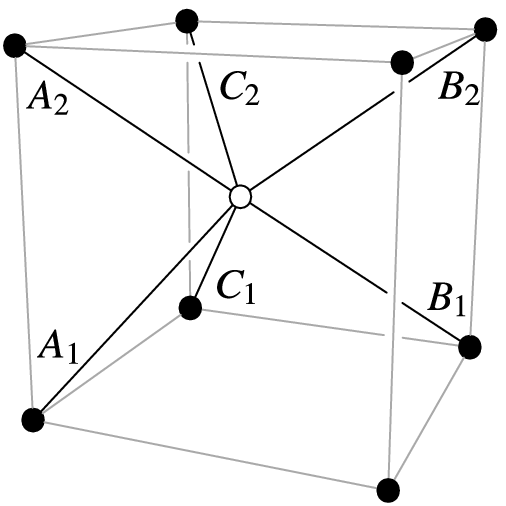}
\end{minipage}
\begin{minipage}[b]{0.40\linewidth}
\raisebox{0.8cm}{
\begin{tabular}{r|cccccccc}
  & $A_1$ & $A_2$ & $B_1$ & $B_2$ & $C_1$ & $C_2$ \cr \hline
$Q_1$ & + & -- & -- & + & 0 & 0 \cr
$Q_2$ & -- & + & 0 & 0 & + & -- 
\end{tabular}
}
\raisebox{1cm}{
\hskip 1cm $W=0$.
}
\end{minipage}
\caption{$D_3$} \label{XXXch}
\end{center}
\end{figure}


\begin{figure}[htbp]
\begin{center}
\begin{minipage}[b]{0.35\linewidth}
\includegraphics[width=4.8cm]{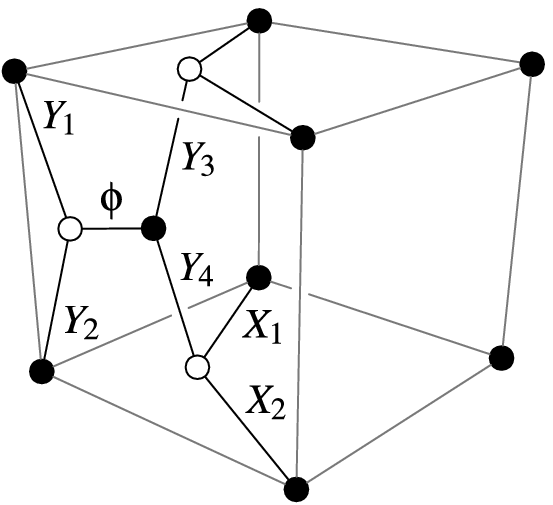}
\end{minipage}
\begin{minipage}[b]{0.55\linewidth}
\raisebox{0.8cm}{
\begin{tabular}{r|ccccccc}
  & $\phi$ & $X_1$ & $X_2$ & $Y_1$ & $Y_2$ & $Y_3$ & $Y_4$ \cr \hline
$Q$ & 0 & 0 & 0 & + & -- & + & --      
\end{tabular}
}
\\
\raisebox{1.5cm}{
$
W = \phi(Y_1 Y_2 - Y_3 Y_4) - X_1 X_2 (Y_1 Y_2 - Y_3 Y_4).
$
}
\end{minipage}
\caption{$\IC^2/\IZ_2 \times \IC^2$} \label{Z2ch}
\end{center}
\end{figure}

\begin{figure}[htbp]
\begin{center}
\begin{minipage}[b]{0.30\linewidth}
\includegraphics[width=4.5cm]{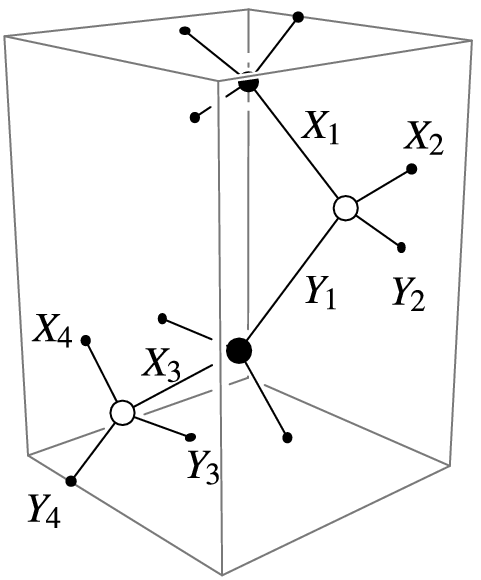}
\end{minipage}
\begin{minipage}[b]{0.60\linewidth}
\raisebox{1.2cm}{
\begin{tabular}{r|cccccccc}
  & $X_1$ & $X_2$ & $X_3$ & $X_4$ & $Y_1$ & $Y_2$ & $Y_3$ & $Y_4$ \cr 
  \hline
$Q_1$ & -- & + & -- & + & 0 & 0 & 0 & 0 \cr
$Q_2$ & 0 & 0 & 0 & 0 & + & -- & + & --  
\end{tabular}
}
\\
\raisebox{1.5cm}{
$
W = X_1 X_2 Y_1 Y_2 - X_3 X_4 Y_1 Y_2 + X_3 X_4 Y_3 Y_4 - X_1 X_2 Y_3 Y_4.
$  
}
\end{minipage}
\caption{$(\IC^2/\IZ_2)^2$} \label{Z22ch}
\end{center}
\begin{center}
\begin{minipage}[b]{0.30\linewidth}
\includegraphics[width=4.5cm]{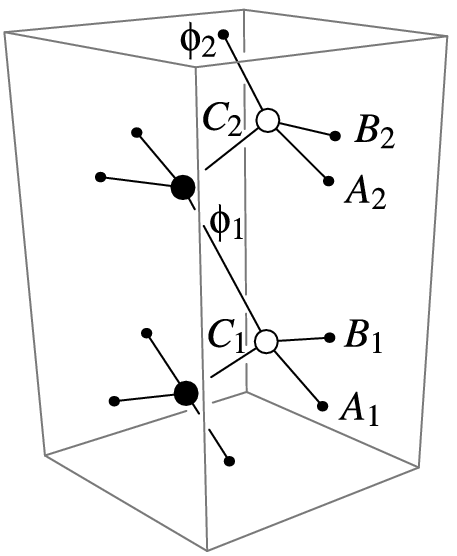}
\end{minipage}
\begin{minipage}[b]{0.60\linewidth}
\raisebox{1.2cm}{
\begin{tabular}{r|cccccccc}
  & $\phi_1$ & $\phi_2$ & $A_1$ & $A_2$ & $B_1$ & $B_2$ & $C_1$ & $C_2$ \cr
\hline
$Q_1$ & 0 & 0 & + & -- & -- & + & 0 & 0 \cr
$Q_2$ & 0 & 0 & -- & + & 0 & 0 & + & -- 
\end{tabular}
}
\\
\raisebox{1.5cm}{
$
W = \phi_1 (A_1 B_1 C_1 - A_2 B_2 C_2) - \phi_2 (A_1 B_1 C_1 - A_2 B_2 C_2).
$
}
\end{minipage}
\caption{$\IC^3/(\IZ_2\times\IZ_2)\times \IC$} \label{Z2Z2ch}
\end{center}
\end{figure}


\begin{figure}[htbp]
\begin{center}
\begin{minipage}[b]{0.35\linewidth}
\includegraphics[width=4.5cm]{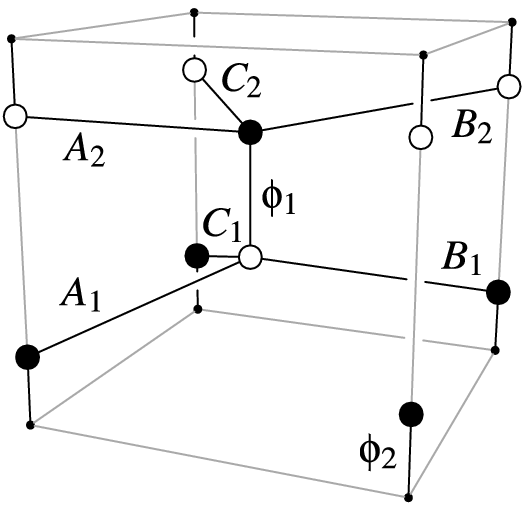}
\end{minipage}
\begin{minipage}[b]{0.60\linewidth}
\raisebox{1.2cm}{
\begin{tabular}{r|cccccccc}
  & $\phi_1$ & $\phi_2$ & $A_1$ & $A_2$ & $B_1$ & $B_2$ & $C_1$ & $C_2$ \cr
\hline
$Q_1$ & 0 & 0 & + & + & -- & -- & 0 & 0 \cr
$Q_2$ & 0 & 0 & -- & -- & 0 & 0 & + & + 
\end{tabular}
}
\\
\raisebox{1cm}{
$
W = \phi_1 (A_1 B_1 C_1 - A_2 B_2 C_2) - \phi_2 (A_1 B_1 C_1 - A_2 B_2 C_2).
$
}
\end{minipage}
\caption{$dP_3 \times \IC$} \label{dP3ch}
\end{center}
\end{figure}

An astute reader may have noticed that the charge matrix of $C(Q^{1,1,1})$ 
and that of $dP_3\times \IC$ are the same except that 
$C(Q^{1,1,1})$ does not have the $\phi_{1,2}$ fields that $dP_3\times \IC$ has. Similarly, $D_3$ and $\IC^3/(\IZ_2\times \IZ_2)$ have almost identical charge 
matrices. We will discuss their implications in section 7.


\section{Moduli Space of Vacua}

Given a toric CY$_4$ and the associated crystal, 
it is natural to expect that 
the abelian theory defined in the previous section  
has a moduli space of vacua, $\CM_V$, that coincides with the CY$_4$. 
We now show that it is indeed the case. 
First, we present a few illustrative examples 
to familiarize the reader with the abelian theory and its $\CM_V$. 
Then we give a general proof that $\mbox{dim}(\CM_V)=4$ 
in parallel with a similar proof in the tiling model. 
Finally, we make a comparison with the D-brane gauge theory 
in the $\CN=4$ cases. 

\subsection{Examples}

\bn

\item
$C(Q^{1,1,1})$ 

We have eight gauge invariant variables $z_{ijk} = A_iB_jC_k$ ($i,j,k=1,2$). 
Label them as
\be
z_0 = A_1B_1C_1, \;\; z_1 = A_1B_1C_2,\;\;
z_2 = A_1B_2C_1, \;\; z_3 = A_1B_2C_2,
\nn \\
z_4 = A_2B_1C_1, \;\; z_5 = A_2B_1C_2,\;\;
z_6 = A_2B_2C_1, \;\; z_7 = A_2B_2C_2.
\ee
They satisfy the quadratic relations 
\be
z_0 z_7 = z_1 z_6 = z_2 z_5 = z_3 z_4, 
\nn \\
z_0 z_3 = z_2 z_1, \;\;\; z_7 z_4 = z_5 z_6, 
\nn \\
z_0 z_5 = z_1 z_4, \;\;\; z_7 z_2 = z_6 z_3, 
\nn \\
z_0 z_6 = z_4 z_2, \;\;\; z_7 z_1 = z_3 z_5.
\ee
As noted earlier in \cite{ot,fab}, this is precisely 
the algebraic definition of $C(Q^{1,1,1})$.

\item 
$\IC^2/\IZ_2 \times \IC^2$ 

In addition to the neutral fields, $\phi$, $X_1$, $X_2$, 
we have four gauge invariant variables: 
$w_1 = Y_1Y_2$, $w_2 = Y_3 Y_4$, $u_1 = Y_1 Y_4$, $u_2 = Y_2 Y_3$. 
The F-term conditions demands that 
\be
\phi = X_1 X_2, \;\;\;\;\; w_1 = w_2 \; (\equiv w).
\ee
We note that $(X_1, X_2)$ span the $\IC^2$ part and the 
algebraic relation, $u_1 u_2 = w^2$, describes $\IC^2/\IZ_2$.

\item 
$(\IC^2/\IZ_2)^2$ 

The $X$ fields and $Y$ fields decouple from each other. 
The F-term condition demands that $X_1 X_2 = X_3 X_4 \; (\equiv w)$.
The gauge-invariant coordinates $u_1 = X_1 X_4$ and 
$u_2 = X_2 X_3$ satisfy $u_1 u_2 = w^2$, which gives $\IC^2/\IZ_2$.
Similarly, the $Y$ fields produce another factor of $\IC^2/\IZ_2$.

\item 
$\IC^3/(\IZ_2\times\IZ_2)\times \IC$

There are seven gauge invariant coordinates,
$\phi_1$, $\phi_2$, 
$w_1 \equiv A_1 B_1 C_1$, $w_2 \equiv A_2 B_2 C_2$\, 
$z_1 \equiv A_1 A_2$, $z_2 \equiv B_1 B_2$, $z_3 \equiv C_1 C_2$,
and two F-term conditions,
\be
\label{z22al}
\phi_1 = \phi_2 \;(\equiv \phi), 
\;\;\;\;\;
w_1 = w_2 \;(\equiv w).
\ee
Clearly, $\phi$ parameterizes the $\IC$ factor and 
the algebraic relation, $z_1 z_2 z_3 = w^2$, 
describes the orbifold $\IC^3/(\IZ_2\times \IZ_2)$.

\item
$dP_3 \times \IC$

The gauge-invariant coordinates are $\phi_1$, $\phi_2$ and $z_{ijk} = A_i B_j C_k$. 
The F-term condition gives two linear constraints:
\be
\label{dp3al}
\phi_1 = \phi_2 \;(\equiv \phi), 
\;\;\;\;\;
z_{111} = z_{222} \;(\equiv w)
\ee 
Label the remaining six $z_{ijk}$ as 
\be
s_1 = A_2 B_1 C_1, && s_2 = A_1 B_2 C_2,
\nn \\
t_1 = A_1 B_2 C_1, && t_2 = A_2 B_1 C_2, 
\nn \\
u_2 = A_1 B_1 C_2, && u_2 = A_2 B_2 C_1. 
\ee
The gauge-invariant coordinates are subject to the quadratic relations, 
\be
w^2 = s_1 s_2 = t_1 t_2 = u_1 u_2 ,
\nn \\
w s_1 = t_2 u_2, \;\;\; w s_2 =t_1 u_1, 
\nn \\
w t_1 = u_2 s_2, \;\;\; w t_2 =u_1 s_1, 
\nn \\
w u_1 = s_2 t_2, \;\;\; w u_2 =s_1 t_1.
\ee
This coincides with the known algebraic definition of $dP_3$. 
As a check, we note that the character function of $dP_3$ \cite{count1},
\[ 
Z=\frac{1+4t+t^2}{(1-t)^3} = 1 + 7 t + 19 t^2 + \CO(t^3),
\]
requires that there be seven monomials of degree one and nineteen 
independent monomials of degree two. 
It is easy to see that we have the correct number of monomials. 
\be
\mbox{deg 1} &:& w, s_i, t_i, u_i 
\nn \\
\mbox{deg 2} &:& w^2 \; ; \; w s_i, w t_i, w u_i \; ; \; s_i^2, t_i^2, u_i^2 \; ; 
\nn \\
&& s_1 t_2, s_2 t_1, t_1 u_2, t_2 u_1, u_1 s_2, u_2 s_1.
\ee
All other monomials of degree two are redundant due 
to the quadratic relations.

\en

\subsection{Dimension counting}

As a warm-up exercise, we first review how to count the dimension of $\CM_V$ 
in the tiling model. Recall the well-known relations in the tiling model:

\bn

\item
$\sharp$(gauge groups) $= \sharp$(faces) $\equiv f$.

\item
$\sharp$(bi-fundamentals) $= \sharp$(edges) $\equiv e$.

\item
$\sharp$(super-potential terms) $= \sharp$(vertices) $\equiv v$. 

\en
To count dim($\CM_V$) of the abelian theory, we note that

\bn
\item 
$\sharp$(F-term conditions) $\equiv  w =v-2$.

The F-term conditions equate all the super-potential terms, 
leading to $v-1$ equations. One of them turns out to be redundant. 

\item 
$\sharp$(D-term conditions) $\equiv q =f-1$.

The D-term conditions come from the $U(1)$ gauge groups. 
All matter fields are neutral under the diagonal $U(1)$, 
since they are all bi-fundamentals. 

\en
The dimension of $\CM_V$ is the number of matter fields 
minus the number of constraints: $\mbox{dim}(\CM_V) = e - w -q$.
Combining the topological condition, $f -e + v =0$,  
with the relations mentioned above,  
we find 
\be
\mbox{dim}(\CM_V) = e - (v-2) - (f-1) = 3, 
\ee
in agreement with the fact that $\CM_V$ coincides with the CY$_3$.

A similar argument can be made for the abelian theory of the crystal model.
The following relations continue to hold.
\bn

\item
$\sharp$(bi-fundamentals) $= \sharp$(edges) $\equiv e$.

\item
$\sharp$(super-potential terms) $= \sharp$(vertices) $\equiv v$. 

\item 
$\sharp$(F-term conditions) $\equiv  w =v-2$.

\item 
$\mbox{dim}(\CM_V) = e - w - q$, where $q$ is the number of D-term conditions.

\en
Now, we introduce as many faces as we need to partition the $T^3$ 
into several cells, so that the topological condition $-c + f - e +v =0$ 
can be used. Here, $c$ and $f$ are the number of cells and faces, respectively. 
Note that, for a given crystal, $f-c = e - v$ is a topological invariant 
that does not depend on the details of the partition. 

It is easy to see that $q$ depends only on the difference $f-c$. 
When a collection of faces surround a cell, the total charge 
vanishes identically, because the matter loops either penetrate 
the faces twice with opposite orientations or do not penetrate them at all. 
Therefore, we can remove faces while keeping $f - c$ fixed 
until we arrive at $c = 1$. Let $\bar{f}$ be the number of faces 
in this configuration. By construction, $\bar{f} -1 = f-c$. 

The remaining $\bar{f}$ faces come in two distinct types. 
The first kind of faces live in the {\em interior} of the cell, 
so that they can be removed without changing $c = 1$. 
All of them give independent $U(1)$ charges. 
The second kind of faces participate in forming the {\em walls} (2-cycles) 
surrounding the unit cell. 
The overall charge corresponding to the sum of all faces on the same wall vanishes, again because the loops either penetrate the faces twice or 
do not penetrate them at all. 
Since three independent walls are needed to make up a three-torus, 
the number of independent gauge groups gets reduced by three. 
Therefore, we have $q = \bar{f} - 3 = f - c - 2$, 
which in turn implies 
\be
\mbox{dim}(\CM_V) = e - (v-2) - (f-c-2) = 4.
\ee 
This is consistent with the fact that $\CM_V$ coincides with the CY$_4$.

\subsection{Comparison with D-brane gauge theory}

It is widely believed that the world-volume theory of M2-branes 
in flat space-time can be obtained by taking the 
strongly coupled, infrared limit of the $\CN=8$ super-Yang-Mills theory 
living on D2-branes. 
The attempt to understand the M2-brane theory from the infrared limit 
of the D2-brane theory was extended to the $\CN=4$, orbifold examples 
in \cite{por}. 
We will now review the result of \cite{por} in the abelian context 
and compare it with our new model. 

The M2-branes probe the product of two singluar 
ALE (asymptotically locally euclidean) spaces. 
Regarding the $U(1)$-fiber 
in one of the ALE spaces as the M-theory circle, the
theories can be considered as gauge theories
on the D2-brane probe in type IIA theory with several D$6$
branes (and possibly O6 planes), transverse to $\IR^3$.

To be concrete, we focus on the $\IC^2/\IZ_n \times
\IC^2/\IZ_k$ examples.  Compactifying on a circle in
$\IC^2/\IZ_k$, we get $k$ D$6$-branes, transverse to
$\IR^3$, with the D2-brane probe. 
The D2-brane and all its mirror images under the $\IZ_n$ orbifold action 
defines a $U(1)^n$ gauge theory. 
Each of the $n$ vector multiplets has three scalars which 
represent the position in the transverse $\IR^3$. 
The hyper-multiplets $X$ are bi-fundamentals 
connecting adjacent $U(1)$-factors of the gauge group \cite{doug}. 
In addition, there are other hyper-multiplets $\phi$
arising from the open string modes connecting the D2- and D6-branes,
which are charged under only one of the $U(1)$ gauge groups.

There are two types of geometric moduli parameters we can add to the
theory. Each D-term equation for the $X$ fields allows a 
Fayet-Iliopoulos (FI) parameter. 
Since all $X$ fields are neutral under the diagonal $U(1)_D$, 
we have $(n-1)$ independent FI parameters $\zeta_i$. 
They correspond to the moduli of the first ALE space, $\IC^2/\IZ_n$. 
One can also introduce the
$k$ mass parameters, $m_i$, for the $\phi$ fields, which originate from  the
relative distance between D2- and D6-branes. 
The total mass $\sum_i m_i$ can be eliminated by shifting
the origin of the Coulomb branch. 
The remaining $(k-1)$ mass parameters can be
understood as the moduli of the second ALE space, $\IC^2/\IZ_k$.

In the $\mathcal{N}=2$ language, the $\CN=4$ 
D-term equations describing the Higgs branch of $X$ fields are given by 
\begin{eqnarray}\label{DtermX}
  |X^{1}_{\ i, i+1}|^2 - |X^{2}_{\ i+1, i}|^2 -
  |X^{1}_{\ i-1,i}|^2 + |X^{2}_{\ i, i-1}|^2 &=& \zeta_i^{D}
  \nonumber \\ X^{1}_{\ i, i+1} X^{2}_{\ i+1, i} - X^{1}_{\
  i-1,i} X^{2}_{\ i, i-1} &=& \zeta_i^{F},
\end{eqnarray}
where $i$ runs over $\{1, \cdots, n-1 \}$. 
The real parameter $\zeta^D$ and complex parameter $\zeta^F$ 
together form a triplet of $\CN=4$ FI parameters. 
The crystal model has only $\CN=2$ manifest supersymmetry, 
and the meaning of $\zeta^F$ is not clear at present. 
The equations for the $\phi$ fields are
\begin{eqnarray}
  \sum_{j=1}^{k-1} \left( |\phi^1_{\ j}|^2 - |\phi^2_{\ j}|^2 \right) =
  0,~~~ \ \ \sum_{j=1}^{k-1} \left( \phi^1_{\ j} \phi^2_{\ j} \right) = 0.
\end{eqnarray}
Note that the equations for $\phi$ are decoupled
from those for $X$ as well as the FI parameters.

Let us now examine $\CM_V$ 
with both the mass and the FI parameters turned on. 
Since the fields $\phi$ become massive, their Higgs branch is lifted.
Turning on the FI terms leads to breaking gauge symmetry down to the
diagonal $U(1)_D$ under which only the fields $\phi$ are charged.
Therefore, $\CM_V$ is the product of
the Higg branch, $\CM_H$, for the $X$ fields described by the D-term equations
(\ref{DtermX}) and the Coulomb branch, $\CM_C$, for the $U(1)_D$ gauge
theory with $k$ massive hyper-multiplets $\phi$. 

As is well known,
the D-terms equations (\ref{DtermX}) coincide with 
Kronheimer's hyper-K\"ahler quotient construction \cite{K} of the first
ALE space, $\IC^2/\IZ_n$, with the singularities resolved by the FI parameters. 
The Higgs branch, $\CM_H$, does not 
receive any quantum corrections, and remains to be 
$\IC^2/\IZ_n$ even in the strong coupling regime. 

The classical Coulomb
branch for the $U(1)_D$ gauge theory is $\IR^3 \times S^1$. 
As in the $\CN=8$ case, the periodic scalar is the Hodge dual of the 
$U(1)_D$ gauge field, and 
the radius of the circle goes to infinity in the IR limit.
The quantum correction 
changes the Coulomb branch significantly \cite{SW96}, so that 
it describes the second ALE space
$\IC^2/\IZ_k$ with the singularities resolved by the mass parameters. 
To summarize, the moduli space of vacua of the full theory is 
$\CM_V = \CM_H \times \CM_C = \IC^2/\IZ_n \times \IC^2/\IZ_k$, 
resolved by the FI and mass parameters.

If we now choose the M-theory circle in $\IC^2/\IZ_n$, 
instead of $\IC^2/\IZ_k$, then we have $n$ D6-branes
with the D2-brane probe. Since the moduli space of vacua, $\widetilde{\CM}_V$, 
of this dual theory should capture the same orbifold geometry, it must be that 
\[
\CM_H = \widetilde{\CM}_C, 
\;\;\;\;\; 
\CM_C = \widetilde{\CM}_H.
\]
It is easy to confirm this from the gauge theory. 
In the dual theory, the first ALE
space $\IC^2/\IZ_n$ is described by the Coulomb branch for
$U(1)$ gauge theory with $n$ massive hyper-multiplets
$\tilde{\phi}$ and the second $\IC^2/\IZ_k$ by the Higgs
branch for $k$ hyper-multiplets $Y$. The duality also exchanges
the mass parameters and the FI parameters. 
The key observation of \cite{por} is that this is an example of the 
mirror symmetry of three dimensional gauge theories \cite{IS96}, 
and that M-theory ``explains'' why mirror symmetry holds in this case.

\begin{figure}[htbp]
\begin{center}
\includegraphics[width=9.0cm]{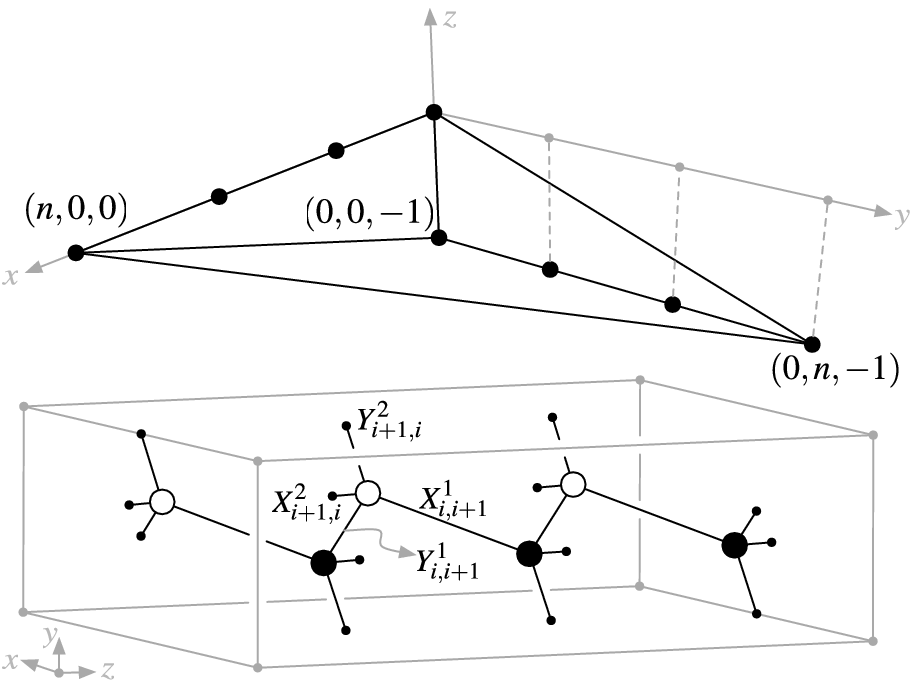}
\caption{Toric diagram and crystal for $(\IC^2/\IZ_n)^2$.} \label{znzn}
\end{center}
\end{figure}

We now turn to our crystal model. 
For simplicity, we restrict our attention to the $k=n$ cases. 
The toric diagram and the crystal for $\IC^2/\IZ_n \times \IC^2/\IZ_n$ 
are depicted in Figure \ref{znzn}. 
The gauge group of the abelian model is factorized as 
$\left(U(1)^n/U(1)_D\right)_X
\times \left(U(1)^n/U(1)_D\right)_Y$. 
The matter fields $(X^1_{\ i,
i+1} , X^2_{\ i+1, i})$ are charged only under the first factor, 
$(U(1)^n/U(1)_D)_X$, as 

\begin{center}
\begin{tabular}{l|cc}
  & $X^1_{i,i+1}$ & $X^2_{i+1,i}$   \cr \hline
$Q_i$ & + & $-$  \cr
$Q_{i+1}$ & $-$ & + 
\end{tabular}
\hskip 1cm 
($i=1, \cdots, n$)
\end{center}
with all other charges vanishing. 
The matter fields
$(Y^1_{\ j, j+1} , Y^2_{\ j+1, j})$ are charged
only under the second factor, $(U(1)^n/U(1)_D)_Y$, in the same way as the 
$X$ fields. 
The super-potential of the crystal model is
\begin{eqnarray}
\label{wzn}
  \mathcal{W} = \sum_{i=1}^{n} \left( X^{1}_{\ i, i+1} X^{2}_{\ i+1, i}
  Y^{1}_{\ i, i+1} Y^{2}_{\ i+1, i} - X^{1}_{\ i-1, i} X^{2}_{\ i, i-1}
  Y^{1}_{\ i, i+1} Y^{2}_{\ i+1, i} \right),
\end{eqnarray}
from which we find the F-term conditions,
\begin{eqnarray}
  X^{1}_{\ i, i+1} X^{2}_{\ i+1, i} - X^{1}_{\
  i-1,i} X^{2}_{\ i, i-1} &=& 0, \nonumber \\
  Y^{1}_{\ j, j+1} Y^{2}_{\ j+1, j} - Y^{1}_{\
  j-1,j} Y^{2}_{\ j, j-1} &=& 0.
\end{eqnarray}
Comparing the field contents, gauge groups and F-term conditions 
with those of the D2-brane gauge theories, 
we note that $\CM_V$ of the crystal model can be regarded as 
the product of the Higgs branch, $\CM_H$, 
of the first description and the Higgs branch, $\widetilde{\CM}_H$, 
of the dual description.
\footnote{Similar Higgs branches also appear in the hyper-K\"ahler quotient 
description of the $d=3$, $\CN=3$ CFT discussed in \cite{billo, ly, hu}.}
In other words, 
\be
\CM_V({\rm crystal}) = \CM_H \times \widetilde{\CM}_H.
\ee
Unlike in the D2-brane gauge theory, where the distinction 
between the Coulomb and Higgs branches is inevitable, 
the crystal model treats the two factors on an equal footing. 

Note that, although the moduli space of vacua factorizes, 
the super-potential (\ref{wzn}) has products of $X$ fields and $Y$ fields. 
It would be interesting to understand how the super-potential (\ref{wzn}) 
arises from the dynamics of the D2-brane gauge theory.

\section{Toric Duality}

It was observed in \cite{crystal2} that $C(T^{1,1})\times\IC$ admits two 
different crystals (See Figure \ref{t11} below). 
The result was based on the fast forward algorithm. 
Let us compare the abelian gauge theories 
of the two crystals.

\begin{figure}[htbp]
\begin{center}
\includegraphics[width=8.0cm]{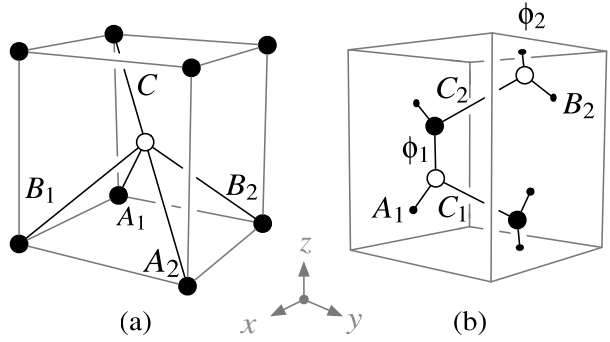}
\caption{Two crystals for $C(T^{1,1})\times \IC$, 
reproduced from \cite{crystal2}.} \label{t11}
\end{center}
\end{figure}

\noindent
The
field theory corresponding to crystal (a) has one independent gauge group. 
The charge assignment for the matter fields is as follows:
\begin{eqnarray}
\label{t11_ch}
\begin{array}{c|ccccc}
 &  A_1 & A_2 & B_1 & B_2 & C  \\
\hline Q & + & + & - & - &0
\end{array} \;\;\; .
\end{eqnarray}
The super-potential vanishes as usual since it has only two atoms. 
The gauge invariant variables parameterizing the moduli space of vacua are
\begin{eqnarray}
z_1 = A_1 B_1,~~~z_2 = A_1 B_2,~~~z_3 = A_2 B_1,~~~z_4 = A_2
B_2,~~~w=C
\end{eqnarray}
which satisfy the relation $z_1 z_3 = z_2 z_4$. The moduli space is clearly
$C(T^{1,1}) \times \bdC$. 

For crystal (b), all matter fields are neutral and the super-potential
takes the form
\begin{eqnarray}
W &=& \phi_1 A_1 C_1  - \phi_2 A_1 C_1 + \phi_2 B_2 C_2 - \phi_1 B_2 C_2,
\end{eqnarray}
from which we derive the F-term conditions:
\begin{eqnarray}
\label{t11f}
A_1 C_1 &=& B_2 C_2,~~~ \phi_1 ~=~ \phi_2.
\end{eqnarray}
The moduli space is again $C(T^{1,1}) \times \bdC$. 

Thus we see that the two crystals give two field
theories describing the same toric singularity. 
The same kind of degeneracy of four-dimensional abelian gauge theories 
was discovered in \cite{he00}, and was named {\bf toric duality}.  
It was shown later that toric duality is realized in the non-abelian quiver 
gauge theories as Seiberg duality. 

To investigate the fascinating possibility of  ``non-abelian'' toric duality in M-theory, 
we will have to construct the ``non-abelian'' theory first, which 
is beyond the scope of this paper. 
In the rest of this section, we will show that, at least in the abelian context, 
toric duality of the crystal model is similar to that of the tiling model 
in many respects.

\subsection{GLSM charge matrix}

In the discussion of toric duality in the tiling model \cite{he00}, 
the gauged linear sigma model (GLSM) description of the 
toric singularity plays a crucial role. 
The matter fields of the GLSM are precisely the perfect matchings 
$p_{\alpha}$. 
The charge matrix $\mathcal{Q}$ of the GLSM can be divided into two parts, 
$\CQ = (\mathcal{Q}_F, \mathcal{Q}_D)^T$, 
where $\CQ_F$ is related to the F-term conditions of the (abelian) 
field theory while $\CQ_D$ originates form the D-term conditions. 
Let us briefly review how to construct $\mathcal{Q}_F$ and $\mathcal{Q}_D$ 
from the field theory. See \cite{he00} for details.  

As we discussed in section 2, the perfect matchings solve the F-term 
conditions automatically via
\begin{eqnarray}
\label{xp2}
X_i = \prod_{\alpha} {p_{\alpha}}^{\langle X_i, p_{\alpha}
\rangle},
\end{eqnarray}
where $\langle X_i, p_{\alpha} \rangle$ equals 1 if $p_{\alpha}$
contains the bond $X_i$ and 0 otherwise.
In general, the number of perfect matchings $\{p_\a\}$ can be larger than 
the number of fields $\{X_i\}$.
Following \cite{he00}, as an intermediate step, 
we can choose a minimal set of independent fields $\{\tilde{X}_m\}$  
after solving the F-term conditions. 
Define a matrix $T$ as
\begin{eqnarray}
T_{m \alpha} = \langle \tilde{X}_m, p_{\alpha} \rangle.
\end{eqnarray}
Then, $\mathcal{Q}_F$ is determined as the cokernel of the $T$, 
\begin{eqnarray}
\sum_{\alpha}T_{m \alpha} {\mathcal{Q}_F}_{\hat{m}\alpha} &=& 0.
\end{eqnarray}
By construction, all fields $X_i$ are neutral under $\CQ_F$.  
They are only charged under the abelian gauge groups of the field theory, 
which are related to $\CQ_D$. Concretely, $\CQ_D$ assigns charges 
$(\mathcal{Q}_D)_{a \alpha}$ to the perfect matchings $p_\a$ such that 
the charges of the abelian gauge theory $Q_{ai}$ is reproduced through (\ref{xp2}).

As a consistency check, let us count the dimension of $\CM_V$ of the GLSM. 
As discussed in section 4, 
the number of independent matter fields $\{\tilde{X}_m\}$ is $e-v+2$ 
and the number of relevant $U(1)$ gauge groups is $q=e-v-2$. 
Suppose that there are $p$ perfect matchings. 
The charge matrix $\mathcalQ$ is then a
$\left( p-(e-v+2)+(e-v-2) \right) \times p = (p-4)\times p$ matrix, which
implies that the dimension of the moduli space of vacua of the 
GLSM is four as expected. 

The GLSM can be used to study blow-up of the toric singularity. 
The blow-ups are controlled by the FI parameters associated to the gauge groups. 
In the tiling/crystal models, since the GLSM charge matrix is constructed 
from the field theory underlying the tiling/crystal, 
not all FI parameters are allowed in the GLSM. 
The $\CQ_D$ part of the GLSM charge matrix can inherit the 
FI parameters from the gauge theory, but 
the $\CQ_F$ part is not allowed to have the FI parameters.

\subsection{ $C(T^{1,1}) \times \bdC $ revisited}

\begin{figure}[htbp]
\begin{center}
\includegraphics[width=11cm]{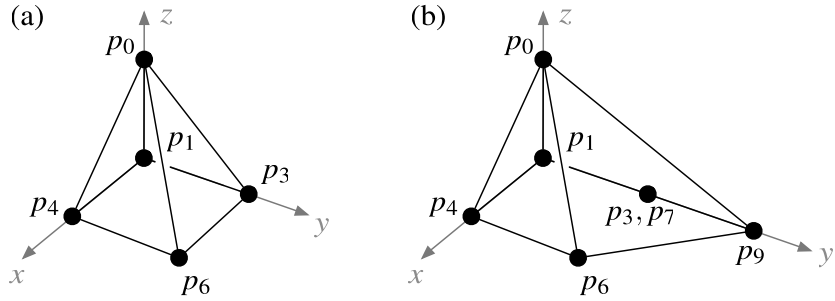}
\caption{Perfect matchings of $C(T^{1,1})\times \IC$ and SPP$\times \IC$.} 
\label{t11pm}
\end{center}
\end{figure}

Let us construct the GLSM charge matrices of the two crystals 
for $C(T^{1,1})\times \IC$.
For crystal (a), all matter fields are independent since there is no
super-potential. In terms of perfect matchings (Figure \ref{t11pm}(a)), 
they are written as
\begin{eqnarray}
A_1=p_1,~~B_2=p_3~~,B_1=p_4,~~A_2 =p_6,~~C=p_0,
\end{eqnarray}
which implies that the matrix $T$ is an identity matrix and that $\mathcal{Q}_F=0$.
Therefore, the GLSM charge matrix consists of only $\mathcal{Q}_D$, 
which is identical to the charge of the abelian gauge theory: 
\begin{eqnarray}
\mathcal{Q} = \begin{pmatrix} 1 & -1 & -1 & 1 & 0 &; \zeta
\end{pmatrix},
\end{eqnarray}
where we included the FI parameter $\zeta$ of the field theory.

For crystal (b),
the independent matter fields are chosen as $\{\phi_1, C_1, A_1, B_2
\}$ after solving the F-term conditions (\ref{t11f}). These matter
fields are written in terms of the perfect matching as
\begin{eqnarray}
\phi_1= p_{0},~~C_1=p_{1} p_{3},~~A_1=p_4 p_6,~~B_2= p_3 p_6 ,
\end{eqnarray}
which leads to
\begin{eqnarray}
T = \begin{pmatrix}  0 & 0 & 0 & 0 & 1 \\ 1 & 1 & 0 & 0 &
  0 \\ 0 & 0 & 1 & 1 & 0 \\ 0 & 1
  & 0 & 1 & 0 \end{pmatrix} \leftarrow \left(
\begin{array}{c|ccccc}
  ~ & p_1 & p_3 & p_4 & p_6 & p_{0}  \\ \hline
  \phi_1 & 0 & 0 & 0 & 0 & 1 \\ 
  C_1 & 1 & 1 & 0 & 0 & 0 \\ 
  A_1 & 0 & 0 & 1 & 1 & 0 \\ 
  B_2 & 0 & 1 & 0 & 1 & 0
\end{array}\right).
\end{eqnarray}
The charge matrix $\mathcal{Q}_F$ (cokernel of $T$) is
\begin{eqnarray}
\mathcal{Q}_F = \begin{pmatrix} 1 & -1 & -1 & 1 & 0\end{pmatrix}.
\end{eqnarray}
The charge matrix $\mathcal{Q}_D$ is absent since all the
matter fields in the theory are neutral. So, the GLSM charge matrix
becomes
\begin{eqnarray}
\mathcal{Q} = \begin{pmatrix} 1 & -1 & -1 & 1 & 0 ; 0\end{pmatrix},
\end{eqnarray}
where the last entry means that the FI term is absent.

We therefore conclude that the GLSM charge matrices of the 
two theories are identical, but their physical origin is 
different. One comes from the D-term conditions, 
and the other from the F-term conditions. 
This phenomenon, called F-D ambiguity in \cite{he00}, 
lies at the heart of toric duality. 
A crucual difference is that an FI term can be introduced in crystal (a), 
but not in crystal (b). In the next section, we will see what difference 
they make when we study possible blow-ups of the singularity. 

\subsection{Another example}

\begin{figure}[htbp]
\begin{center}
\includegraphics[width=10.0cm]{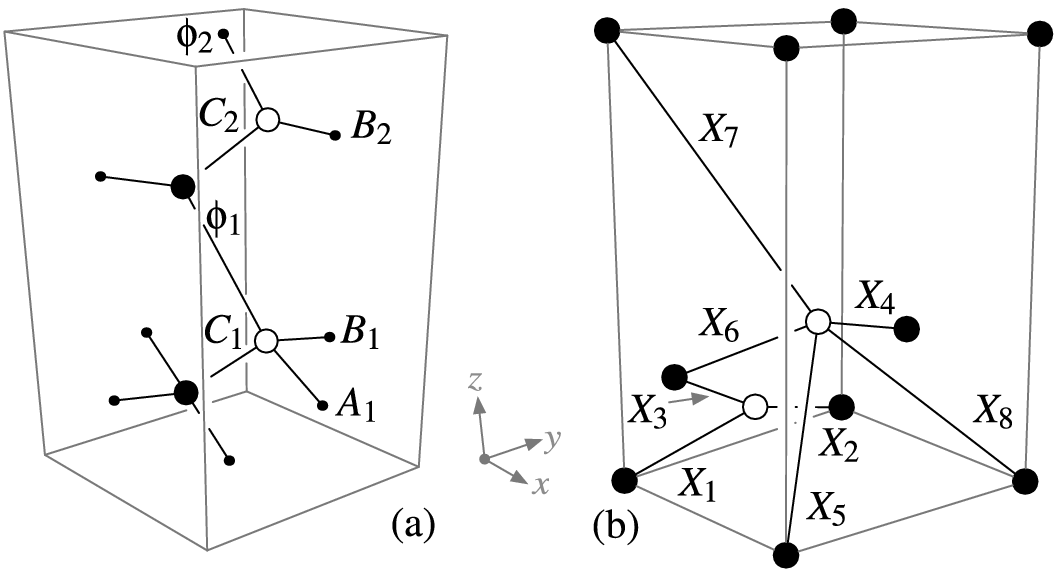}
\caption{Two crystals for SPP$\times\IC$.} \label{spp}
\end{center}
\end{figure}

We consider another illustrative example of toric duality, namely, 
$SPP \times \bdC$. We have again two different crystals depicted in
Figure \ref{spp}. 
For crystal (a), the field theory has seven matter fields
whose charges under the relevant $U(1)$ gauge group are assigned
as
\begin{eqnarray}
\begin{array}{c|ccccccc}
& \phi_1 & \phi_2 & A_1 & B_1 & B_2 & C_1 & C_2 \\ \hline 
Q & 0 & 0 & 0 & + & - & - & +
\end{array}.
\end{eqnarray}
The super-potential $W$ takes the form
\begin{eqnarray}
W &=& 
\phi_1 A_1 B_1 C_1 - \phi_2 A_1 B_1 C_1 + \phi_2 B_2 C_2 - \phi_2 B_2 C_2,
\end{eqnarray}
which gives the F-term conditions
\begin{eqnarray}
A_1 B_1 C_1 = B_2 C_2,~~\phi_1 = \phi_2.
\end{eqnarray}
After solving these F-term conditions, we choose the independent
matter fields to be 
$\{ \phi_1, C_1, A_1, B_1, C_2 \}$. They are
written in terms of perfect matchings (Figure \ref{t11pm}(b)) as
\begin{eqnarray}
\phi_1=p_{0},~~C_1=p_1 p_3,~~A_1=p_4 p_6,~~B_1= p_7 p_9,~~C_2=p_1
p_4 p_7,
\end{eqnarray}
from which we find
\begin{eqnarray}
T = \begin{pmatrix} 0 & 0 & 0 & 0 & 0 & 0 & 1 \\ 1 & 1 & 0 & 0 & 0
& 0 & 0 \\ 0 & 0 & 1 & 1 & 0 & 0 & 0 \\ 0 & 0 & 0 & 0 & 1 & 1 & 0
\\ 1 & 0 & 1 & 0 & 1 & 0 & 0\end{pmatrix}
\;\;\;\;\; \mbox{and} \;\;\;\;\;
%
%
\mathcal{Q}_F = \begin{pmatrix} p_1 & p_3 &p_4 &p_6 & p_7 &p_9 &
  p_0 \\ 1 & -1 & 0 & 0 & -1 & 1 & 0
\\ 1 & -1 & -1 & 1 & 0 & 0 & 0\end{pmatrix}.
\end{eqnarray}
The gauge charges of matter fields are reproduced if we choose 
$\mathcal{Q}_D$ to be
\begin{eqnarray}
\mathcal{Q}_D = \begin{pmatrix} 0 & -1 & 0 & 0 & 1 & 0 &
0\end{pmatrix}.
\end{eqnarray}
Altogether, the GLSM charge matrix $\mathcal{Q}$ for the crystal (a) is given
by
\begin{eqnarray}
\label{sppa}
\mathcal{Q} = \begin{pmatrix} 1 & -1 & 0 & 0 & -1 & 1 & 0 &;0
\\ 1 & -1 & -1 & 1 & 0 & 0 & 0 &;0\\ 0 & -1 & 0 & 0 & 1 & 0 &
0 &;\zeta\end{pmatrix}.
\end{eqnarray}

For crystal (b), the charges of eight matter fields for the abelian theory are
\begin{eqnarray}
\label{sppb_qq}
\begin{array}{r|cccccccc}
~ & X_1 & X_2 & X_3 & X_4 & X_5 & X_6 & X_7 & X_8 \\ \hline 
Q_1 & 0 & 0 & 0 & - & - & + & 0 & + \\
Q_2 & - & + & 0 & - & 0 & + & 0 & 0
\end{array}.
\end{eqnarray}
We can also read off the super-potential from the four atoms in the
crystal,
\begin{eqnarray}
W = X_1 X_2 X_3 + X_4 X_5 X_6 X_7 X_8- X_3 X_4 X_6 - X_1 X_2
X_5 X_7 X_8,
\end{eqnarray}
which gives the F-term conditions
\begin{eqnarray}
X_5 X_7 X_8 = X_3~~, X_1 X_2 = X_4 X_6.
\end{eqnarray}
With gauge invariant monomials 
\be
z_1=X_5 X_8, &  z_2=X_2 X_4 X_8, & z_3=X_1 X_5 X_6, \nn \\ 
z_4=X_3, &  z_5=X_7, & w=X_1 X_2(=X_4 X_6), 
\ee
the moduli space of vacua can be described as
\begin{eqnarray}
z_1 w^2 = z_2 z_3,~~ z_1 z_5 = z_4,
\end{eqnarray}
which is an algebraic description of $SPP \times \bdC$ singularity. 
In terms of perfect matchings, the matter fields are written by
\begin{eqnarray}\label{perfectnewSPP}
X_1=p_1 p_3,~~X_2=p_7 p_9,~~X_3=p_4 p_6 p_{0},~~~~~ \nonumber
\\ X_4=p_3 p_9,\ X_5=p_4~~,X_6=p_1 p_7,~~X_7=p_{0},~~X_8=p_6.
\end{eqnarray}
By the method introduced in the previous subsection, 
we can construct the GLSM charge matrix $\mathcal{Q}$
\begin{eqnarray}
\label{sppb}
\mathcal{Q} = \begin{pmatrix} p_1 & p_3 & p_4 & p_6 & p_7 & p_9 & p_0 &
  \\ 1 & -1 & 0 & 0 & -1 & 1 & 0 &;0~
\\ 1 & -1 & -1 & 1 & 0 & 0 & 0 &;\zeta_1\\ 0 & -1 & 0 & 0 & 1 & 0 &
0 &;\zeta_2\end{pmatrix},
\end{eqnarray}
which is the same as (\ref{sppa}) except that the
second row in this case comes from a D-term rather than an F-term.
These two crystals are therefore toric dual.

\section{Partial resolution}

As we have seen in previous section,
the crystal model encodes the geometry of the CY singularity 
via K\"ahler quotient (GLSM). 
We can take K\"ahler deformations of the singularity by
introducing the blow-up parameters $\zeta_a$ to the moment maps. 
The moment maps with generic parameters would resolve the singularity
completely. 
For special values of the parameters, some residual singularities may survive. 
This procedure is known as partial resolutions.

Partial resolutions in the context of AdS$_5$/CFT$_4$ has been extensively 
discussed, for example, in \cite{mp}.
Let us recall how D-branes ``know'' about the smoothing.
When a D3-brane probes the CY singularity, 
the closed string modes in the twisted sector 
couple to open string modes as the FI terms on the
world-volume gauge theory. 
Turning on the FI terms induces vevs for some of the
charged matter fields, 
which leads to (partial) gauge symmetry breaking via Higgs mechanism. 
Keeping light fields only, one obtains the low energy effective 
theory which describe the D3-brane sitting at the residual singularity. 

Partial resolution in the tiling model was studied most systematically in \cite{ur1}.
In the tiling diagram, partial resolutions results in taking off
the edges corresponding to matter fields which acquire the vacuum
expectation value. It results in another tiling diagram consistent
with the remaining ``daughter'' singularity. 
The most efficient method of determining which edge to remove 
utilize the amoeba projection and the perfect matchings \cite{ur1}. 
Here, we will use the older approach of \cite{he00} 
based on the GLSM of perfect matchings, since it shows 
the connection to the abelian gauge theory more clearly. 
Let us briefly summarize the procedure.

We begin by choosing a vertex of the toric diagram we would like to delete. 
The vertex should be at a corner of the toric diagram; otherwise, the 
remaining toric diagram will not be convex. Then we consider 
the moment map (D-term) equation of the GLSM with 
all allowed FI parameters included. 
The M-theory origin of the FI parameters is not clear, 
but we proceed with the assumption that they exist.
We look for a solution to the moment map 
equation with a non-zero vev for the perfect matching 
sitting on the vertex to be deleted. 
The relation between matter fields $\{X_i\}$ of the gauge theory 
and the perfect matchings $p_\a$ of the GLSM then determines 
which bonds of the crystal should be removed by the partial resolution. 
Let us work out several illustrative examples in detail.

\subsection{$\IC^3/(\IZ_2 \times \IZ_2)\times \IC\goto {\rm SPP}\times \IC$} 

We first consider the field theory for the crystal 
$\IC^3/(\IZ_2 \times \IZ_2)\times \IC$. See Figure 7, where
the matter fields with their charges and the super-potential are given. 
The perfect matchings $p_{\alpha}$ of this crystal are shown in 
Figure \ref{z2z2mat} below. 

\begin{figure}[htbp]
\begin{center}
\includegraphics[width=8cm]{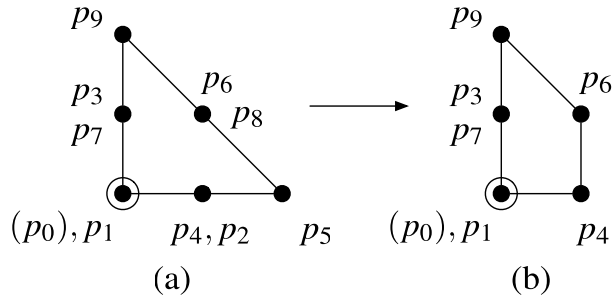}
\caption{(a) Perfect matchings of $\IC^3/(\IZ_2\times \IZ_2)\times\IC$ 
projected onto the $(x,y)$-plane. 
All perfect matchings except $p_0$ lie on the plane. 
(b) Perfect matchings of SPP$\times \IC$ after partial resolution, 
which is identical to Figure 17(b).} 
\label{z2z2mat}
\end{center}
\end{figure}

\noindent
In terms of the perfect matchings $p_{\alpha}$, the matter fields are
written as
\begin{eqnarray}
A_1=p_4 p_5 p_6,~ B_1=p_7 p_8 p_9, ~ C_1=p_1 p_2 p_3, ~\phi_1=p_{0} ~\nonumber \\
A_2 = p_2 p_5 p_8,~ B_2=p_3 p_6 p_9, ~C_2=p_1 p_4
p_7,~\phi_2=p_{0}.
\end{eqnarray}
Following the procedure of subsection 5.1, we obtain the GLSM charge matrix, 
\begin{eqnarray}
\mathcal{Q}(\mathbb{C}^3/(\IZ_2 \times \IZ_2)\times \mathbb{C}) &=&
\left(%
\begin{array}{ccccccccccl}
p_1 & p_2 & p_3 & p_4 & p_5 & p_6 &
p_7 & p_8 & p_9 & p_{0} & ~  \\ 1 & 0 & -1 & 0 & 0 & 0 & -1 & 0 & 1 & 0 & ;0 \\
1 & -1 & 0 & 0 & 0 & 0 & -1 & 1 & 0 & 0 & ;0 \\ 1 & 0 & -1 & -1 &
0 & 1 & 0 & 0 & 0 & 0 & ;0\\ 1 & -1 & 0 & -1 & 1 & 0 & 0 & 0 & 0 & 0 & ;0\\
-1 & 1 & 1 & 0 & 0 & -1 & 0 & 0 & 0 & 0 & ;\zeta_1\\ 0 & 0 & -1 &
0 & 0 & 0 & 1 & 0 & 0 & 0 & ;\zeta_2
\end{array}
\right).
\end{eqnarray}
Compared with the GLSM charge matrix of $d=4$, $\mathcal{N}=1$ 
gauge theory on a D3-brane probing 
$\IC^3/(\IZ_2 \times \IZ_2)$,
the crucial difference is that our present model can have at most 
two FI parameters, whereas the D3-brane theory has three.
In what follows, we will show that due to the lack of 
FI parameters, the M2-brane theory cannot capture all possible 
partial resolutions allowed in the D3-brane theory. 

The toric diagram of $\IC^3/(\IZ_2\times \IZ_2)\times\IC$ in Figure \ref{z2z2mat} 
suggests that we can remove a point, say $p_5$, to obtain a partially resolved 
singularity SPP$\times \IC$. Let us check whether such a partial resolution 
can be realized in the GLSM and the abelian theory 
with a suitable choice of the FI terms. 

In the GLSM, we should solve the corresponding moment map equation,
\begin{eqnarray}
\mathcal{Q} ~\vec{y} = (0, 0, 0, 0, \zeta_1, \zeta_2)^{\text{T}} ,
\end{eqnarray}
where $y_{\alpha}$ denotes $|p_{\alpha}|^2$. 
Since the fields of the GLSM are the perfect matchings, 
we know what to expect of the partial resolution. 
We want to remove $p_5$ from the toric diagram, so it should get a vev.
On the other hand, the perfect matchings unaffected by the resolution,  
$\{ p_1 , p_3, p_7, p_9 \}$, should not get vevs. 
The unique solution satisfying these requirements is 
\begin{eqnarray}
\vec{y} = (0, \zeta_1, 0, 0, \zeta_1, 0, 0, \zeta_1, 0, 0)^{\text{T}},
\end{eqnarray}
with $\zeta_2=0$. We find that, in addition to $p_5$, 
$p_2$ and $p_8$ also get vevs.
To summarize, we obtain the toric diagram for SPP$\times \IC$ 
by removing $p_2, p_5, p_8$; see Figure \ref{z2z2mat}(a,b).

Let us translate these results into the
field theory language. 
The D-term equations with the FI terms are
\footnote{
We use the same notation for the FI parameters of the GLSM and 
those of the abelian gauge theory, but in general 
their numerical values are not the same.}
\begin{eqnarray}
|C_1|^2 - |C_2|^2 - |A_1|^2 + |A_2|^2 &=& \zeta_1  \nonumber
\\ |B_1|^2 -
|B_2|^2 - |C_1|^2 + |C_2|^2 &=& \zeta_2.
\end{eqnarray}
When we turn on FI term $\zeta_1$ only,
the point we choose on the moduli space of vacua is represented by
the vev
\begin{eqnarray}
A_2 = {\zeta_1}^{1/2}
\end{eqnarray}
with the others vanishing, since the matter field $A_2$ is written as
products only of resolved perfect matchings $\{p_2, p_5, p_8\}$.
After integrating out massive modes, the low energy theory is governed by 
the super-potential
\begin{eqnarray}
W= \phi_1 (A_1 B_1 C_1 - B_2 C_2 ) - \phi_2 (A_1 B_1 C_1 -
B_2 C_2 ),
\end{eqnarray}
with one relevant $U(1)$ gauge symmetry under which the matter
fields are charged as
\begin{eqnarray}
\begin{array}{c|ccccccc}
 & A_1 & B_1 & C_1 & \phi_1 & B_2 & C_2 & \phi_2 \\ \hline 
 Q & 0 & + & - & 0 & - & + & 0
\end{array} ~.
\end{eqnarray}
This theory is nothing but the gauge theory for 
crystal (a) in Figure \ref{spp}. 
Note that this crystal can be obtained simply by eliminating
the bond $A_2$ in the crystal for $\mathbb{C}^3/(\IZ_2 \times
\IZ_2)\times \mathbb{C}$; compare Figures \ref{Z2Z2ch} and \ref{spp}(a).

\subsection{${\rm SPP}\times\IC \goto C(T^{1,1})\times \IC$ {\rm or} 
$\IC^2/\IZ_2 \times \IC^2$}

We now ask whether we can further resolve 
SPP$\times \mathbb{C}$ to reach the daughter singularities 
$C(T^{1,1})\times\IC$ or $\IC^2/\IZ_2 \times \IC^2$. 
Recall that in subsection 5.3, we analyzed the field theories 
for the two crystals for SPP$\times \IC$ that are toric dual to each other. 
We analyze both of them in turn. 

For crystal (a) in Figure \ref{spp}, let us summarize the field theory 
information here for convenience. We can express 
the matter fields in terms of perfect matchings  as
\begin{eqnarray}
\label{sppmat}
A_1=p_4 p_6,~ B_1=p_7 p_9, ~ C_1=p_1 p_3, ~\phi_1=p_{0} ~\nonumber \\
B_2=p_3 p_6 p_9, ~C_2=p_1 p_4 p_7,~\phi_2=p_{0},
\end{eqnarray}
and the GLSM charge matrix is given by
\begin{eqnarray}
\label{sppa_ch}
\mathcal{Q}(\text{SPP}\times \mathbb{C}) &=&
\left(%
\begin{array}{cccccccl}
p_1 & p_3 & p_4 & p_6 &
p_7 & p_9 & p_{0} & ~  \\ 1 & -1 & 0 & 0 & -1 & 1 & 0 & ;0 \\
1 & -1 & -1 & 1 & 0 & 0 & 0 &;0 \\ 0 & -1 & 0 & 0 & 1 & 0 & 0 &
;\zeta
\end{array}
\right).
\end{eqnarray}
In the GLSM, we begin again by solving the moment map equation,
\begin{eqnarray}
\label{sppmom}
\mathcal{Q} ~\vec{y} = (0, 0, \zeta)^{\text{T}} .
\end{eqnarray}
To obtain $\mathcal{C}({T^{1,1}}) \times \mathbb{C}$, 
we require that $p_9$ get a vev. It is easy to find a solution,
\begin{eqnarray}
\vec{y} = (0, 0, 0, 0, \zeta, \zeta, 0)^{\text T},
\end{eqnarray}
implying that $p_7$ also gets a vev.  This leads to a toric
diagram for $\mathcal{C}({T^{1,1}}) \times \mathbb{C}$ 
with the surviving perfect matchings $\{ p_0, p_1, p_3, p_4, p_6\}$.
On the gauge theory side, the D-term equation is

\begin{eqnarray}
|B_1|^2 - |B_2|^2 - |C_1|^2 + |C_2|^2 &=& \zeta.
\end{eqnarray}
{}From (\ref{sppmat}), we see that 
$B_1 = p_7 p_9$ is the only field to acquire a vev,
\begin{eqnarray}
B_1={\zeta}^{1/2}.
\end{eqnarray}
The low energy theory becomes the gauge theory for crystal (b) in Figure \ref{t11}. 
As a check, note that this can be obtained from crystal (a) 
in Figure \ref{spp} by taking off the bond $B_1$. 
\footnote{To avoid confusion, 
note that the coordinate axes are oriented differently in the two figures.}

Next, we try to obtain $\IC^2/\IZ_2 \times \IZ_2$ by giving a vev to $p_6$. 
However, a solution to (\ref{sppmom}) with a non-zero vev for $p_6$ 
does not exist. All other solutions of (\ref{sppmom}) correspond 
to the complete resolution of the singularity or 
the partial resolution to $C(T^{1,1})\times \IC$. 
Thus we see that unlike the D3-brane theory, the M2-brane theory 
probing $\IC^3/(\IZ_2\times \IZ_2) \times \IC$, 
does not have room for enough FI parameters to 
allow partial resolution to all conceivable daughter singularities.

We now turn to study partial resolutions of crystal (b) in Figure \ref{spp}.
We again summarize the field theory information here for convenience.
The GLSM charge matrix is
\begin{eqnarray}
\label{sppb_ch}
\tilde{\mathcal{Q}}(\text{SPP}\times \IC) &=&
\left(%
\begin{array}{cccccccl}
p_1 & p_3 & p_4 & p_6 &
p_7 & p_9 & p_{0} & ~  \\ 1 & -1 & 0 & 0 & -1 & 1 & 0 & ;0 \\
1 & -1 & -1 & 1 & 0 & 0 & 0 &;\zeta_1 \\ 0 & -1 & 0 & 0 &
1 & 0 & 0 & ;\zeta_2
\end{array}
\right).
\end{eqnarray}
Compared to (\ref{sppa_ch}), we have an additional FI parameter. 
We will see that it opens up new possibilities for partial resolutions. 
The D-term equations become
\begin{eqnarray}
-|X_3|^2 - |X_4|^2 - |X_5|^2 + |X_6|^2 + |X_8|^2 &=&
\zeta_1 \nonumber
\\ -|X_1|^2 + |X_2|^2 - |X_4|^2 + |X_6|^2 &=& \zeta_2.
\end{eqnarray}
The super-potential for this theory takes the form
\begin{eqnarray}\label{spnewSPP}
W = X_1 X_2 X_3 + X_4 X_5 X_6 X_7 X_8- X_3 X_4 X_6 - X_1 X_2
X_5 X_7 X_8.
\end{eqnarray}
In terms of perfect matchings, the matter fields can be written as 
\begin{eqnarray}\label{perfectnewSPP2}
X_1=p_1 p_3,~~X_2=p_7 p_9,~~X_3=p_4 p_6 p_{0},~~~~~ \nonumber
\\ X_4=p_3 p_9,\ X_5=p_4~~,X_6=p_1 p_7,~~X_7=p_{0},~~X_8=p_6. \nn
\end{eqnarray}
Solving the moment map equations with the GLSM charge matrix 
(\ref{sppb_ch}), 
we find {\em two} solutions. The first one is
\begin{eqnarray}
\vec{y} = (0, -\zeta_1, 0, 0, 0, -\zeta_2, 0)^{\text T},
\end{eqnarray}
with $\zeta_1=\zeta_2$ ($\neq 0$). 
Since $p_3, p_9$ are resolved, only the matter field $X_4$ acquires a vev
\begin{eqnarray}
X_4 = ({-\zeta_1})^{1/2},
\end{eqnarray}
Replacing $X_4$ by its vev in (\ref{spnewSPP}), we find that 
$X_3$ and $X_6$ get F-term masses. 
In fact, integrating out the massive modes make the super-potential 
vanish completely. 
The remaining theory with one gauge group and five fields 
$\{X_1, X_2, X_5, X_7, X_8\}$ is identical to that of the crystal 
$C(T^{1,1})\times \IC$ in Figure \ref{t11}(a), 
once we relabel the fields by 
\be
X_1 \goto A_1, \;\;\; 
X_2 \goto B_2, \;\;\; 
X_5 \goto B_1, \;\;\; 
X_7 \goto C, \;\;\; 
X_8 \goto A_2. 
\ee
The charges $Q$ in (\ref{t11_ch}) is related to $Q_1$ and $Q_2$ in 
(\ref{sppb_qq}) by $Q=Q_1-Q_2$.

Pictorially, elimination of the bond corresponding to $X_4$ 
produces a bi-valent atom which can be regarded as a mass term. 
Integrating out the massive modes translates into shrinking the 
bi-valent atom and the bonds attached to it. 
The remaining crystal coincides with Figure \ref{t11}(a). 

The other solution of the moment map equation is
given by
\begin{eqnarray}
\vec{y} = (0, 0, 0, \zeta_1, 0, 0, 0)^{\text T}, 
\;\;\;\;\; \zeta_2 = 0.
\end{eqnarray}
In the toric diagram, removing $p_6$ leads to the daughter singularity
$\mathbb{C}^2/\IZ_2 \times \IC^2$. We can check it at the level of the 
field theory. 
{}From (\ref{perfectnewSPP2}), we see that $X_8$ is the only field to 
acquire a vev through $p_6$.
\begin{eqnarray}
X_8 = {\zeta_1}^{1/2}
\end{eqnarray}
The low energy theory around this vacuum
is now governed by the super-potential read off from (\ref{spnewSPP})
\begin{eqnarray}
W = X_1 X_2 X_3 + X_4 X_5 X_6 X_7 - X_3 X_4 X_6 - X_1 X_2 X_5
X_7,
\end{eqnarray}
The charges under the remaining gauge group are given by
\begin{eqnarray}
\begin{array}{c|ccccccc}
 & X_1 & X_2 & X_3 & X_4 & X_5 & X_6 & X_7  \\ \hline  
Q & - & + & 0 & - & 0 & + & 0
\end{array}~.
\end{eqnarray}
This agrees precisely with
the field theory for the crystal $\mathbb{C}^2/\IZ_2 \times \IC^2$ 
summarized in Figure \ref{Z2ch}, if we relabel the fields by 
\be
X_3 \goto \phi, \;\; 
X_1 \goto Y_2, \;\;
X_2 \goto Y_1, \;\;
X_4 \goto Y_4, \;\;
X_6 \goto Y_3, \;\;
X_5 \goto X_1, \;\;
X_7 \goto X_2.  
\ee
One can also verify that removing the bond $X_8$ from the crystal (b) 
in Figure \ref{spp} and changing the basis for the unit cell of the crystal give
the crystal in Figure \ref{Z2ch}.

\begin{figure}[htbp]
\begin{center}
\includegraphics[width=11.0cm]{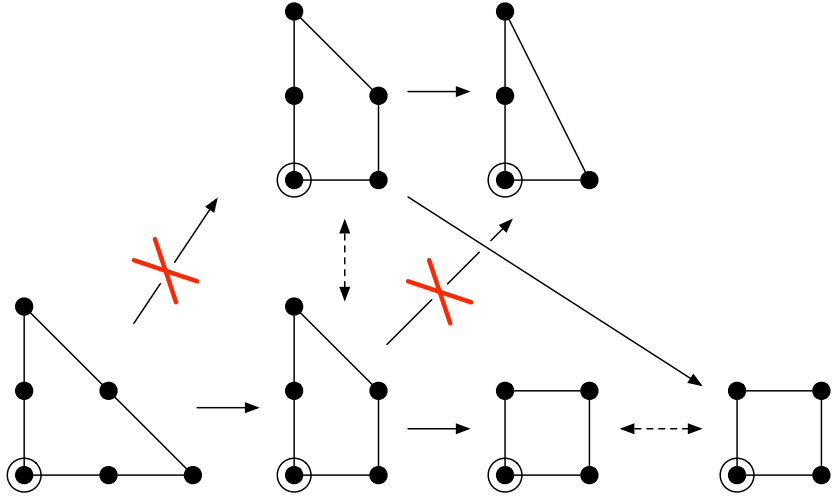}
\caption{Partial resolutions and toric dualities.} \label{resol}
\end{center}
\end{figure}

Figure \ref{resol} summarizes all the results we have obtained so far. 
The solid arrows denote partial resolutions and the dashed bi-directional 
arrows denote toric duality. 

\section{New RG Flows?} 

In AdS$_5$/CFT$_4$, there are two famous holographic RG flows. 
One is the Klebanov-Witten (KW) flow \cite{kw} 
from the orbifold $\IC^2/\IZ_2\times \IC^2$ 
to the conifold $C(T^{1,1})$. The other one is the Pilch-Warner (PW) flow 
\cite{pw1, pw2},
$\CN=4 \goto \CN=1^*$, and orbifolds thereof.
To our knowledge, nearly all 
examples of RG flows in  AdS$_4$/CFT$_3$ discussed in the literature 
\cite{mrg1, mrg2, mrg3, mrg4, mrg5, mrg6, mrg7} 
are analogous to the PW 
flow rather than the KW flow.\footnote{
An exception is Ref. \cite{ot}, where it was 
argued that an RG flow of the KW type 
from $(\IC^2/\IZ_2)^2$ to $C(Q^{1,1,1})$ exist. 
It does not seem to be allowed in our crystal model.} 
We will now use our crystal model to   
argue that new RG flows of the KW type may exist.

\subsection{Klebanov-Witten flow revisited}

\begin{figure}[htbp]
\begin{center}
\includegraphics[width=14.0cm]{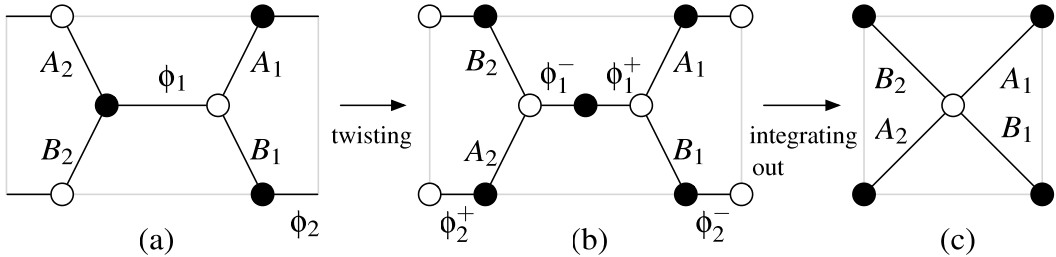}
\caption{Klebanov-Witten flow in the tiling model.} \label{KW}
\end{center}
\end{figure}

We begin with a short review of the KW flow 
with emphasis on its tiling model interpretation.
The super-potential of the $\IC^2/\IZ_2 \times \IC$ 
orbifold is encoded in the tiling (a) in Figure \ref{KW}.
\be
W_{\text{UV}} 
= \Tr( \phi_1 A_1 B_1 - \phi_2 B_1 A_1 + \phi_2 B_2 A_2 - \phi_1 A_2 B_2).
\ee
The RG-flow (massive deformation) is triggered by 
the so-called twisted mass term, 
\be
\label{twim}
\Delta W_{{\rm UV}} = \frac{m}{2}\Tr( \phi_1^2 - \phi_2^2)
\ee
Integrating out $\phi_1$ and $\phi_2$ leaves the super-potential 
\be
\label{wir}
W_{\text{IR}} = \frac{1}{m}\Tr(A_1 B_1 A_2 B_2 - B_1 A_1 B_2 A_2).
\ee
Can we understand not only the end-points of the flow but the entire flow 
in the tiling model? 
The mass term (\ref{twim}) causes a problem, 
because in the tiling model a matter field can appear in a super-potential 
term at most once. 
We can circumvent this difficulty by considering a slightly different UV 
theory which flows to the same theory in the IR. 
The new UV theory, described by the tiling (b) in Figure \ref{KW}, 
has the super-potential,
\be
\widetilde{W}_{\text{UV}} = 
\Tr(\phi_1^+ A_1 B_1 - \phi_2^- B_1 A_1 + \phi_2^+ B_2 A_2 - \phi_1^- A_2 B_2) 
- \Tr (\phi_1^+ \phi_1^- - \phi_2^+ \phi_2^-) .
\ee
Integrating out $\phi_1^\pm$ and $ \phi_2^\pm$ leads to the 
same super-potential as (\ref{wir}) in the IR.

Although the new UV theory is not quite the same as the original 
mass-deformed theory, it is useful for a few reasons. 
First, not only the theory itself, but also the process of integrating out 
massive fields can be described in the tiling model. 
It corresponds to removing the mass terms (bi-valent atoms) by shrinking 
the bonds attached to it; see Figure \ref{KW}(b, c). 
Second, it can be derived from the original, 
undeformed, theory in a simple and systematic way. 

Recall that the inverse algorithm of the tiling model produces 
the tiling graph from the set of intersecting 1-cycles (zig-zag paths).
The transition from tiling (a) to (b) in Figure \ref{KW} can 
be interpreted as an extra twisting of the 1-cycles. See Figure \ref{twist} below. 
Note that the twisting produces a new vertex corresponding to the mass term, 
flips the colors of the vertices on the left half, 
and exchanges the locations of $A_2$ and $B_2$. 

\begin{figure}[htbp]
\begin{center}
\includegraphics[width=11.0cm]{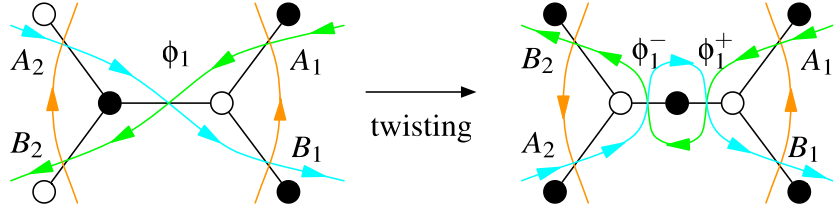}
\caption{Twisting.} \label{twist}
\end{center}
\end{figure}

To conclude,
given the tiling graph for the UV theory, 
the twisting interpretation offers an intuitive and efficient way 
to find out the tiling graph for the IR theory.

\subsection{M-theory flows}

We now study massive deformation of 
the abelian gauge theory in the crystal model. 
We find two examples which resemble the KW flow.
The twisting picture we discussed in the previous subsection again 
gives an intuitive picture of the flow. 
Whether these flows really exist in the  ``non-abelian'' theory 
is a very interesting and important problem. Although a direct 
analysis of the ``non-abelian'' theory is not feasible at present, 
its AdS/CFT dual description on the supergravity side should be possible 
in a way similar to the analysis of the KW flow \cite{pwkw}.
A detailed study of the supergravity flows will be reported elsewhere \cite{rg}.

Our first example is the flow from the  orbifold 
$\IC^3/(\IZ_2\times \IZ_2)\times \IC$ to $D_3$.
We recall from section 3 that the 
$\IC^3/(\IZ_2\times \IZ_2) \times \IC$ model has the charge assignment,
\begin{center}
\begin{tabular}{r|cccccccc}
  & $\phi_1$ & $\phi_2$ & $A_1$ & $A_2$ & $B_1$ & $B_2$ & $C_1$ & $C_2$ \cr
\hline
$Q_1$ & 0 & 0 & + & -- & -- & + & 0 & 0 \cr
$Q_2$ & 0 & 0 & -- & + & 0 & 0 & + & -- 
\end{tabular}
\end{center}
and the super-potential
\be
W = \phi_1 (A_1 B_1 C_1 - A_2 B_2 C_2) - \phi_2 (A_1 B_1 C_1 - A_2 B_2 C_2).
\ee
If we add the twisted mass term, 
\be
\Delta W = \frac{m}{2}( \phi_1^2 - \phi_2^2)
\ee
and integrate out $\phi_1$ and $\phi_2$, 
then the resulting theory has a vanishing super-potential, while 
the charges of the remaining fields remain unchanged. 
This is nothing but the abelian theory of $D_3$.

Let us check how the moduli space of vacua changes. Recall that 
the $\IC^3/(\IZ_2\times \IZ_2)\times \IC$ orbifold 
has the algebraic description 
$z_1 z_2 z_3 = w^2$ with an unconstrained variable $\phi$; 
see the paragraph containing (\ref{z22al}). 
After the RG flow, the variable $\phi$ disappear and 
the algebraic equation is deformed to $z_1 z_2 z_3 = w_1 w_2$. 

\begin{figure}[htbp]
\begin{center}
\includegraphics[width=14.0cm]{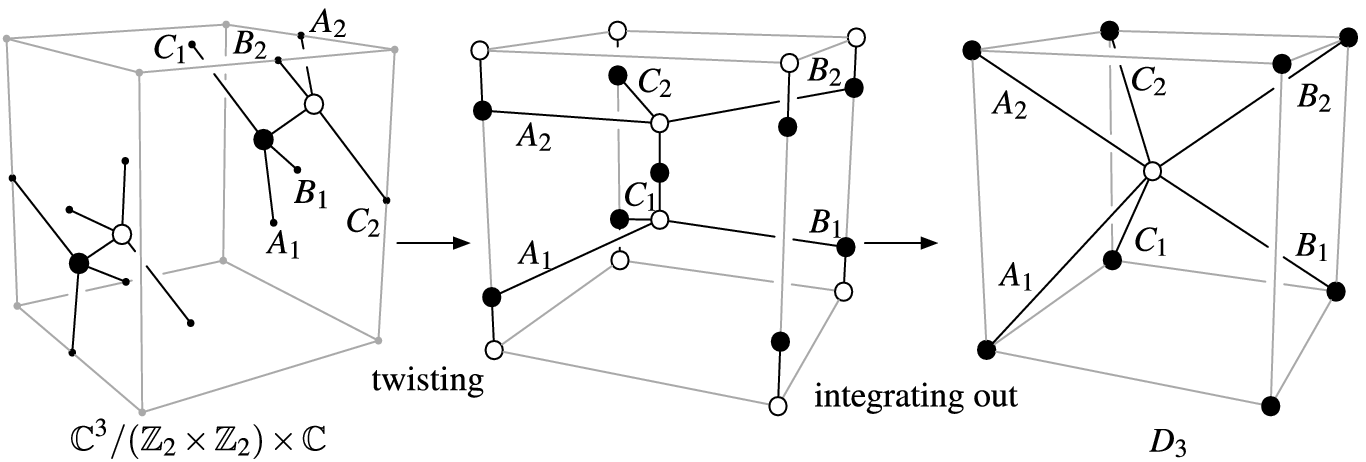}
\caption{RG flow from $\IC^3/(\IZ_2\times \IZ_2)\times \IC$ to $D_3$.}\label{RGX}
\end{center}
\end{figure}

We can summarize the massive deformation in the crystal picture as we 
did for the KW flow; see Figure \ref{RGX}. 
Adding the twisted mass term translates to 
extra twistings on the $\phi_{1,2}$ bonds. 
Unlike in the KW flow, the lattice vector should be changed 
in order to keep the new crystal in a unit cell.

\begin{figure}[htbp]
\begin{center}
\includegraphics[width=14.0cm]{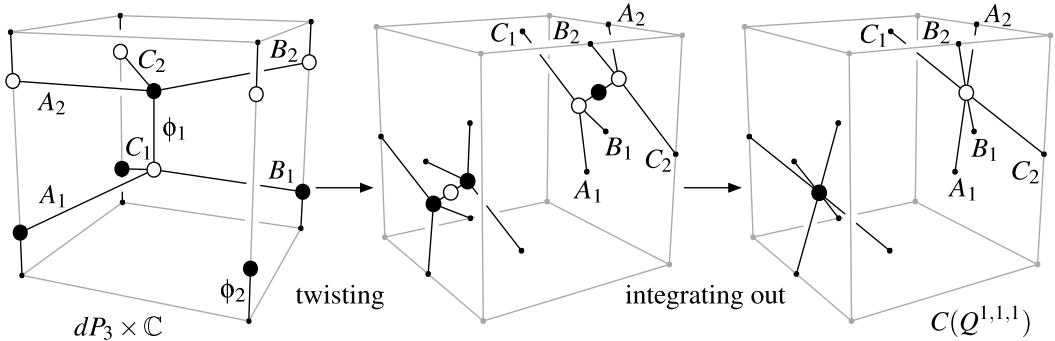}
\caption{RG flow from $dP_3 \times \IC$ to $C(Q^{1,1,1})$.} \label{RGQ}
\end{center}
\end{figure}

The same story holds for our second example,
$dP_3 \times \IC \goto C(Q^{1,1,1})$. 
Recall from section 3 that the $dP_3 \times \IC$ theory has the charge assignment,
\begin{center}
\begin{tabular}{r|cccccccc}
  & $\phi_1$ & $\phi_2$ & $A_1$ & $A_2$ & $B_1$ & $B_2$ & $C_1$ & $C_2$ \cr
  \hline
$Q_1$ & 0 & 0 & + & + & -- & -- & 0 & 0 \cr
$Q_2$ & 0 & 0 & -- & -- & 0 & 0 & + & + 
\end{tabular}
\end{center}

\noindent
and the super-potential
\be
W= \phi_1 (A_1 B_1 C_1 - A_2 B_2 C_2) - \phi_2 (A_1 B_1 C_1 - A_2 B_2 C_2).
\ee
Adding the twisted mass term for $\phi_1$ and $\phi_2$ 
and integrating them out, 
we find a vanishing super-potential and the charge matrix of the 
$C(Q^{1,1,1})$ theory. 
It is also straightforward to show that 
the algebraic equations describing the moduli space of vacua 
are related; see examples 1 and 5 in section 4.1.
The flow is summarized in Figure \ref{RGQ}. 

\section{Discussion}

We made an attempt to write down the world-volume theory 
of a single M2-brane probes in the context of the crystal model. 
The resulting effective theory turned out to be an abelian gauge theory. 
In the crystal model, we identified
the gauge groups and found how to read off the charges of the matter fields. 
The probe theory was shown to have 
the moduli space of vacua which coincides precisely with 
the CY$_4$ associated to the crystal. 
We also discussed toric duality and partial resolution of the probe theory. 
Finally, we found a hint for the existence of new RG flows in
M-theory. 

One interesting observation is that the geometry probed by M2-brane
looks different
from that probed by D-branes and fundamental strings. It is known that
fundamental string can probe geometric as well as non-geometric phases 
while D-branes can probe only geometric phases \cite{Greene, Witten}. 
The study of the
abelian gauge theory of M2-brane suggests that the phase structure is
more restricted. In the abelian gauge theory, this comes from the less
available FI parameters. This also makes the pattern of toric dualities 
different from that observed in D3-brane theories defined on
CY$_3$ singularities. It is an interesting problem to
figure out how such restrictions of moduli space arise. In the abelian gauge theory 
we study, we explored the mesonic branches of the underlying theory and it would be
an interesting problem to find the tools to explore the baryonic
branches. Recent works \cite{bar1, bar2, bar3} may be relevant. 

Aside from the ambitious task of constructing the ``non-abelian'' theory, 
there are a few directions that deserve further study. 
First, we hope to find an explicit description of the special
Lagrangian 
manifold $\S$ and the merged world-volume $\CM$ discussed in section 2 and 3, 
so that we can complete the inverse algorithm and verify 
the behavior of the 2-cycles which give the charges.   

Exactly marginal deformations of the CFT$_3$ is another important topic.  
In AdS$_5$/CFT$_4$, it is known \cite{ima2} that a generic toric CY$_3$ 
with $d$ vertices on the toric diagram admits 
($d-1$) exactly marginal deformations.  
One of them is the diagonal (complexified) gauge coupling. 
Another one is the so-called $\b$-deformation, which 
attaches phases $e^{\pm i\pi \b}$ to the super-potential terms. 
The other $(d-3)$ deformations in the gauge theory 
are combinations of relative gauge couplings and super-potential terms. 
They are interpreted as turning on $B_{NS}+iC_{RR}$
along the $(d-3)$ 2-cycles of $Y_5$; as shown in \cite{t3},  
$H_2(Y_5, \IZ) = \IZ^{d-3}$, where $Y_5$ is the base of the CY$_3$ cone. 
In the crystal model, there is no analog of the overall gauge coupling. 
Nor is there deformation due to a vev of a field along homology cycles, 
because M-theory has only the 3-form field $C$ and  
$H_3(Y_7, \IZ)=0$ generically. So, the $\b$-deformation seems to be 
the only generic marginal deformation. 
On the CFT side, it again attaches phases $e^{\pm i\pi \b}$ to the 
super-potential terms. On the supergravity side, it corresponds 
to turning on the $C$-field along 
the $T^2 \subset T^3$ orthogonal to the $R$-symmetry direction 
\cite{lunin, ahn2, glmw}.
There is a slight puzzle here. In view of the $D=3$, $\CN=2$ supersymmetry, 
the parameter $\b$ is expected to be a complex number, but all the known supergravity solutions have only real values of $\b$. 
It will be nice to resolve this puzzle. It will be also interesting 
to understand non-generic deformations in theories with $\CN=4$ or 
more symmetry using the crystal model and other approaches such as 
\cite{kol}. 

We conclude the discussion with more speculative comments. 
Very recently \cite{gai}, a large class of CFT$_3$ was constructed 
using Chern-Simons theory coupled to matter fields. It will be interesting 
to find out whether there is any overlap between this construction and our crystal model. Finally, the 2d dimer (tiling) model is known to have many deep 
connections with other areas of physics such as the Ising model, mirror symmetry 
of CY$_3$ and black-holes. See \cite{dij} and references therein for more information. 
It is tempting to suspect that similar relation may exist for the 3d dimer (crystal) 
model.

\vskip 0.5cm

\centerline{\bf Acknowledgments}

\vskip .5cm

It is our pleasure to thank Hoil Kim and Ho-Ung Yee for useful discussions. 
Seok Kim is supported in part by the KOSEF Grant R010-2003-000-10391-0.
Sangmin Lee is supported by the KOSEF Basic Research Program, Grant R01-2006-000-10965-0.
Sungjay Lee is supported in part by the Korea Research Foundation Grant R14-2003-012-01001-0.
Jaemo Park is supported by the Science Research Center
Program of KOSEF through the Center for Quantum Space-Time (CQUeST) of
Sogang University with the grant number R11-2005-021.

\end{document}